\documentclass{aa}
\usepackage{graphics,epsfig,txfonts}
\usepackage{longtable}
\usepackage{lscape}
\usepackage{array}

\usepackage{natbib}
\bibpunct{(}{)}{;}{a}{}{,}

\newcommand{\ergcm}[1]{$\times 10^{#1}$ erg cm$^{-2}$ s$^{-1}$}

\newcommand{\ergs}[1]{$\times 10^{#1}$ erg s$^{-1}$}
\newcommand{\oergs}[1]{$10^{#1}$ erg s$^{-1}$}

\newcommand{\ohcm}[1]{$10^{#1}$ cm$^{-2}$}
\newcommand{\expo}[1]{$\times 10^{#1}$}

\newcommand{\fmax}{\hbox{F$_{\rm max}$}}
\newcommand{\fmin}{\hbox{F$_{\rm min}$}}

\newcommand{\ct}{cts s$^{-1}$}

\newcommand{\Halpha}{H${\alpha}$}
\newcommand{\Porb}{P$_{\rm orb}$}
\newcommand{\ltsima}{$\buildrel < \over \sim$}
\newcommand{\lsim}{\lower.5ex\hbox{\ltsima}}
\newcommand{\gtsima}{$\buildrel > \over \sim$}
\newcommand{\gsim}{\lower.5ex\hbox{\gtsima}}
\newcommand{\ang}{\AA}
\newcommand{\xmm}{XMM-{\it Newton}}
\newcommand{\cxo}{{\it Chandra}}

\usepackage{color}

\begin{document}
 
\title{High-mass X-ray binaries in the Small Magellanic Cloud
       \thanks{The catalogue is available at the CDS via anonymous ftp to
               cdsarc.u-strasbg.fr (130.79.128.5) or via
               http://cdsarc.u-strasbg.fr/viz-bin/qcat?J/A+A/???/A??
               and a living edition at http://www.mpe.mpg.de/heg/SMC
              }
      }

\author{      F.~Haberl\inst{1} 
       \and   R.~Sturm\inst{1}  
       }

\titlerunning{High-mass X-ray binaries in the SMC}
\authorrunning{Haberl et al.}

\institute{ Max-Planck-Institut f\"ur extraterrestrische Physik, Giessenbachstra{\ss}e, 85748 Garching, Germany
          }

\date{Received 8 September 2015 / Accepted 28 October 2015}

 \abstract{}
          {The last comprehensive catalogue of high-mass X-ray binaries in the Small Magellanic Cloud (SMC) was published about ten years ago.
           Since then new such systems were discovered, mainly by X-ray observations with \cxo\ and \xmm. For the majority of the proposed HMXBs
           in the SMC no X-ray pulsations were discovered as yet, and unless other properties of the X-ray source and/or the optical counterpart 
           confirm their HMXB nature, they remain only candidate HMXBs.} 
          {From a literature search we collected a catalogue of 148 confirmed and candidate HMXBs in the SMC and investigated their 
          properties to shed light on their real nature. 
          Based on the sample of well-established HMXBs (the pulsars), we investigated which observed properties are most appropriate 
          for a reliable classification.
          We defined different levels of confidence for a genuine HMXB based on spectral and temporal characteristics of the X-ray sources 
          and colour-magnitude diagrams from the optical to the 
          infrared of their likely counterparts. We also took the uncertainty in the X-ray position into account.}
          {We identify 27 objects that probably are misidentified because they lack an infrared excess of the proposed counterpart.
          They were mainly X-ray sources with a large positional uncertainty. This is supported by additional information obtained 
          from more recent observations. Our catalogue comprises 121 relatively high-confidence HMXBs (the vast majority with Be 
          companion stars). About half of the objects show X-ray pulsations, while for the rest no pulsations are known as yet. 
          A comparison of the two subsamples suggests that long pulse periods in 
          excess of a few 100\,s are expected for the ``non-pulsars'', which are most likely undetected because of aperiodic variability 
          on similar timescales and insufficiently long X-ray observations. 
          The highest X-ray variability together with the lowest observed minimum fluxes for 
          short-period pulsars indicate that in addition to the eccentricity of the orbit, its inclination against the plane of 
          the Be star circum-stellar disc plays a major role in determining the outburst behaviour.}
          {The large population of HMXBs in the SMC, in particular Be X-ray binaries, provides the largest homogeneous sample 
          of such systems for statistical population studies.}

 
\maketitle
 
\section{Introduction}
\label{sec:introduction}

High-mass X-ray binaries (HMXBs) are comprised of an early-type star and a compact object that orbit each other. 
The compact object is in most cases a neutron star (NS), but it can also be a black hole 
\citep[see e.g. for the Magellanic Clouds][]{2005A&A...442.1135L} or a white dwarf \citep[][and references therein]{2012A&A...537A..76S}. 
Many of the HMXBs show pulsations in their X-ray flux, which indicate the spin period of the NS.
The Small Magellanic Cloud (SMC) is peculiar because it hosts exceptionally many known HMXBs. 
So far, the optical counterpart is only in one (well-confirmed) case a super-giant star (SMC\,X-1), while 
for all other identified cases a Be star (with Balmer emission lines) was found, forming a Be/X-ray binary (BeXRB). 
The last comprehensive catalogue of HMXBs in the SMC was published by \citet{2005A&A...442.1135L}. 
In the meantime, new multi-wavelength data were collected, and many new objects were found with \cxo, RXTE, 
Swift, and \xmm\ observations of the SMC, which can at least be treated as candidates for this class of X-ray binaries.  
Most of the multi-wavelength work on the BeXRBs in the SMC concentrated on these pulsars \citep{2015MNRAS.452..969C}. 
In this work we present a complete list of known HMXBs in the SMC, including all sources that have been proposed at least as candidates in the literature. 
We use multi-wavelength information to identify some of the candidates as likely mis-identifications,
and in our final catalogue we devise a scheme according to which the objects are a genuine HMXB system
with different confidence levels.

\section{Catalogue}
\label{sec:catalogue}

For an updated catalogue of HMXBs and candidates in the SMC (and the Magellanic Bridge, which extends beyond the 
Eastern Wing) we compiled lists from the literature that present properties from large samples or searches 
for such systems (see Table~\ref{tab:catref}) and include the more recent discoveries of individual objects. 
We collected information about the X-ray sources as well as their companion stars from the literature 
and included it in the catalogue to use it for statistical studies.
An excerpt of our final catalogue is presented in Table~\ref{tab:smccat} of the online material 
(first the 63 pulsars sorted according to pulse period, followed by the other objects sorted by coordinates), 
and the full version will be published as an electronic version. The content is described in Table~\ref{tab:catdesc}.
In the comment column we provide key references for each source, which are listed in Table~\ref{tab:keyref} 
of the online material.

\begin{table*}
\caption[]{Literature for HMXBs in the SMC.}
\begin{center}
\begin{tabular}{ll}
\hline\hline\noalign{\smallskip}
\multicolumn{1}{l}{Reference} &
\multicolumn{1}{l}{Description} \\
\noalign{\smallskip}\hline\noalign{\smallskip}
{\citet{2005A&A...442.1135L}} & HMXB catalogue of the Magellanic Clouds\\ 
{\citet{2005MNRAS.356..502C}} & Optical properties of HMXBs in the SMC\\
{\citet{2005MNRAS.362..879S}} & HMXBs in archival \xmm\ data\\
{\citet{2008MNRAS.383..330M}} & Chandra SMC Wing survey\\
{\citet{2008ApJS..177..189G}} & RXTE observations of SMC pulsars\\
{\citet{2008MNRAS.388.1198M}} & Optical spectroscopy of BeXRB pulsars in the SMC\\
{\citet{2008A&A...489..327H}} & New BeXRBs from \xmm\ observations in 2006 and 2007\\ 
{\citet{2009ApJ...697.1695A}} & Optical identification of Chandra sources\\
{\citet{2009ApJ...707.1080A}} & Optical spectroscopy of BeXRBs in the SMC\\
{\citet{2010ApJ...716.1217L}} & Catalogue of SMC sources from deep Chandra observations.\tablefootmark{a}\\
{\citet{2011MNRAS.413.1600R}} & Long-term optical variability of HMXB pulsars in the SMC\\
{\citet{2013A&A...558A...3S}} & \xmm\ survey of the SMC\\ 
{\citet{2015MNRAS.452..969C}} & BeXRB pulsars in the SMC\\
\noalign{\smallskip}\hline
\end{tabular}
\tablefoot{
\tablefoottext{a}{Includes detections down to two net counts. Most of these low-significance sources 
                  are not found in the Chandra Source Catalogue \citep[CSC][]{2010ApJS..189...37E}}.
}
\end{center}
\label{tab:catref}
\end{table*}

\newcolumntype{L}[1]{>{\raggedright\let\newline\\\arraybackslash\hspace{0pt}}p{#1}}
\begin{table*}
\caption[]{Catalogue description.}
\begin{center}
\begin{tabular}{c L{16.5cm}}
\hline\hline\noalign{\smallskip}
\multicolumn{1}{c}{Column} &
\multicolumn{1}{l}{Description} \\
\noalign{\smallskip}\hline\noalign{\smallskip}
  1        & Source number\\
  2 - 3    & X-ray coordinates, right ascension and declination (epoch 2000.0)\\
  4        & Uncertainty of X-ray position [\arcsec]. For XMM-Newton positions taken from {\citet{2013A&A...558A...3S}} the 1$\sigma$ 
             error includes a systematic uncertainty of 0.5\arcsec.\\
  5        & Origin of the X-ray coordinate (A: ASCA, C: Chandra, E: Einstein, I: Integral, N: \xmm, R: ROSAT, S: Swift, X: RXTE). 
             When no reliable position could be determined from the non-imaging RXTE collimator-instruments, a radius of 30\arcmin\ for 
             the position error indicates the size of the field of view.\\
  6        & Reference for source discovery.\\
  7        & Identification of optical counterpart with emission-line star from {\citet{1993A&AS..102..451M}}. The negative number 
             indicates a star found in the catalogue of \citet{2000MNRAS.311..741M}.\\
  8 - 15   & Flags indicating different source properties. For their description see Table~\ref{tab:flags}.\\
  16       & Confidence class (values 1-6, see Table~\ref{tab:classes}).\\
  17 - 18  & Optical coordinates, right ascension and declination (epoch 2000.0) for the identified counterpart 
             from \citet{2002AJ....123..855Z}, or - when not available there - from \citet{2002ApJS..141...81M}.\\
  19 - 26  & The Magellanic Clouds Photometric Survey (MCPS): U, error(U), B, error(B), V, error(V), I, error(I) [mag] from \citet{2002AJ....123..855Z}.\\
  27 - 32  & Colour indices U-B, error(U-B), B-V, error(B-V), V-I, error(V-I) [mag] derived from MCPS photometry.\\
  33 - 34  & Reddening-free Q-value (Q = U-B-0.72$\times$(B-V)) and error(Q) [mag].\\
  35       & Near-IR counterpart to the optical star from the Two Micron All Sky Survey \citep[2MASS,][]{2006AJ....131.1163S}.\\
  36 - 41  & Near-IR magnitudes with corresponding errors: J, error(J), H, error(H), K, error(K) [mag].\\
  42 - 45  & Near-IR colour indices, J-H, error(J-H), H-K, error(H-K) [mag].\\
  46 - 53  & Spitzer IRAC fluxes at 3.6, 4.5, 5.8 and 8.0 $\mu$m [mag] with respective errors \citep[from the SAGE project, for a description see][]{2006AJ....132.2268M}.\\
  54       & Angular distance between X-ray and optical position [\arcsec].\\
  55       & Angular distance between optical and near-IR position [\arcsec].\\
  56       & Neutron star spin period [s] inferred from X-rays.\\
  57       & Orbital period [days] (see flags for origin).\\
  58 - 59  & Maximum and minimum X-ray flux [erg cm$^{-2}$ s$^{-1}$] when available in the 0.2$-$10 keV band. 
             Fluxes in the SMC \xmm\ catalogue of {\citet{2013A&A...558A...3S}} are given for the 0.2 to 4.5 keV band. To convert 
             them into the 0.2$-$10 keV band, we multiplied them by a factor of 2.6 assuming a standard power law with photon index 
             0.9 {\citep{2008A&A...489..327H}} and a column density of \ohcm{21} (solar abundance). 
             For Swift XRT count rates we used a flux conversion factor of 1.1\expo{-10} erg cm$^{-2}$ cts$^{-1}$\\  
  60       & Flag for minimum flux: 1 for a non-detection with an upper limit; -1 when unknown; 0 for detection.\\
  61 - 62  & References for maximum and minimum X-ray flux.\\
  63       & X-ray variability factor (ratio of maximum to minimum flux).\\
  64       & Equivalent width of the \Halpha\ line [\ang] (minimum value if more than one measurement is available).\\
  65       & Maximum equivalent width of the \Halpha\ line [\ang].\\
  66       & References for the \Halpha\ measurements.\\
  67       & Comments with key references.\\
 
\noalign{\smallskip}\hline
\end{tabular}
\end{center}
\label{tab:catdesc}
\end{table*}

\subsection{Special notes on catalogue sources}

The published lists of HMXBs in the SMC disagree in the details for some entries. 
In particular, sources with uncertain X-ray position or with a pulse period 
detected with low significance can lead to different interpretations of their nature. 
The X-ray position of the majority of sources detected with Chandra or \xmm\ is sufficiently accurate to uniquely identify their optical 
counterpart. However, there are cases whose angular separation exceeds 2-3$\sigma$ confidence, which cause the identification to be uncertain
and increase the probability for a chance coincidence of the X-ray source with an early-type star.
Moreover, the spin periods of pulsars evolve with time, and detections with an uncertain position of the X-ray source may not be associated 
uniquely with an object.
In the following sections we provide information for such cases to introduce our classifications.
In Sect.~\ref{sect:nocat} we present a list of candidates that were previously rejected as HMXB and that we did not include 
in the catalogue.

\subsubsection{Update on individual BeXRB pulsars}
\label{sect:pulsar_notes}

SXP7.92 - A new pulsar with a period of 7.92\,s was discovered by \citep{2008ATel.1600....1C} in RXTE data.
\citet{2009MNRAS.394.2191C} reported six detections of SXP7.92 with RXTE and suggested AzV285 as the probable optical counterpart.
They detected an X-ray source using Swift consistently in position with AzV285.
\citet{2013ATel.5552....1I} detected 7.92\,s pulsations from a source at RA = 00:57:58.4 and Dec = $-$72:22:29.5 (error 1.5\arcsec) 
in Chandra data. The Chandra source is 20.4\arcmin\ away from AzV285, but is fully compatible with the RXTE pulsar.
The source was probably also detected by ROSAT \citep[RX\,J0057.9$-$7222, entry 75 in the HRI catalogue of ][]{2000A&AS..147...75S}.
We conclude that SXP7.92 = CXOU\,J005758.4$-$722229 = RX\,J0057.9$-$7222 is a BeXRB pulsar and the correct optical counterpart is
2MASS\,J00575856$-$7222290. This star shows outbursts every 40.03 days in OGLE II data, which is most likely the orbital period of 
the binary system \citep{2013ATel.5556....1S}. The detection of X-rays from the position of AzV285 with Swift suggests that 
this is the optical counterpart of another BeXRB, which is also confirmed by \xmm\ observations (see below for XMM\,J010155.7$-$723236).

SXP9.13 - Pulsations with a period of 9.13\,s were discovered from the ASCA source AX\,J0049$-$732 \citep{1998IAUC.7040....2I}.
The source is located in a crowded region of the SMC bar with two hard ROSAT sources (RX\,J0049.5$-$7310 and RX\,J0049.2$-$7311) as 
possible counterparts \citep{2000A&A...361..823F}. RX\,J0049.5$-$7310 was confirmed to be a BeXRB with a spin period of 894 s 
\citep{2010ApJ...716.1217L}. Consequently, several authors have
alloted RX\,J0049.2$-$7311 to the 9.13 s pulsar. However, 
RX\,J0049.2$-$7311 was covered by \xmm\ and \cxo\ observations many times with sufficient photon statistics to expect a 
detection of the 9.13 s period. This was never seen, and therefore we do not identify RX\,J0049.2$-$7311 as the 9.13 s 
pulsar AX\,J0049$-$732, but instead keep two separate BeXRBs \citep[within the ASCA error circle of 40$\arcsec$  are $\sim$25 possible 
counterparts brighter than V = 18 mag with $-0.2 <$ B-V $< 0.2$ in the OGLE BVI photometric catalogue;][]{1998AcA....48..147U}.

SXP82.4 - XTE\,J0052$-$725 was discovered as a new transient in RXTE data \citep{2002IAUC.7932....2C}, and X-ray pulsations were found 
in archival Chandra observations. From the RXTE SMC monitoring \citet{2008ApJS..177..189G} identified an outburst pattern with a 
period of 362.3$\pm$4.1 d in X-rays. \citet{2011MNRAS.413.1600R} analysed the OGLE III data, which revealed a significant peak at 
171$\pm$0.3 d in the power spectra, which is less than half the reported X-ray period. \citet{2008ApJS..177..189G} already noted
that the X-ray period is longer than would be expected given its spin period position in the Corbet diagram 
\citep[][see also Fig.~\ref{fig:porb_spin}, in which SXP82.4 is found as the right-most open red circle]{1984A&A...141...91C,2005ApJS..161...96L}.
We add here that the long X-ray period also places XTE\,J0052-725 outside the relation between \Porb\ and the equivalent width 
of the \Halpha\ line 
\citep[][see also Fig.~\ref{fig:porb_Ha}, in which SXP82.4 is again found as the right-most open red circle]{1997A&A...322..193R}. 
This suggests that the orbital period might be closer to the period found in the optical.

SXP91.1 - Pulsations with $\sim$91\,s were found in RXTE and ASCA data \citep[XTE\,J0053$-$724 and AX\,J0051$-$722,][]{1998IAUC.6803....1C}. 
RXTE detections of pulsations at 85.4\,s and 89.0\,s were initially assigned to individual pulsars, but were later recognised 
as most likely stemming from one pulsar with a high spin-down rate \citep{2008ApJS..177..189G,2010ATel.2813....1C,2011ATel.3396....1C}.
For our catalogue we only considered one pulsar (SXP91.1).

\begin{figure}
  \resizebox{\hsize}{!}{\includegraphics[clip=]{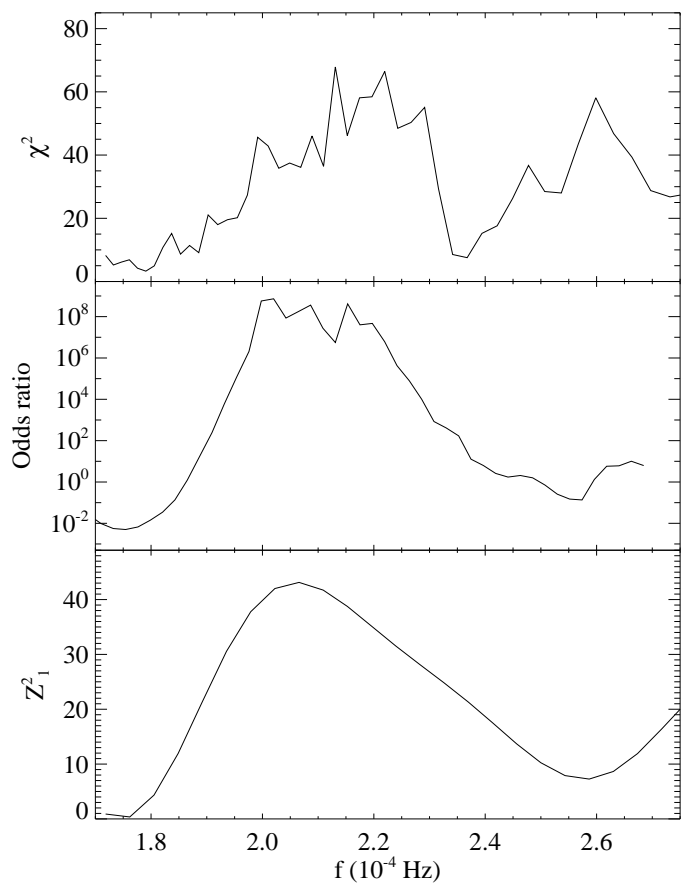}}
 \caption{
    Periodograms obtained from the combined EPIC data of SXP4693 from the \xmm\ observation 0601210801 (0.2$-$10 keV). $\chi^2$ test, 
    Bayesian odds ratio, and Rayleigh Z$^2_1$ test are shown from top to bottom.
  }
  \label{fig:sxp4693_timing}
\end{figure}

\begin{figure}
  \resizebox{\hsize}{!}{\includegraphics[clip=,angle=-90]{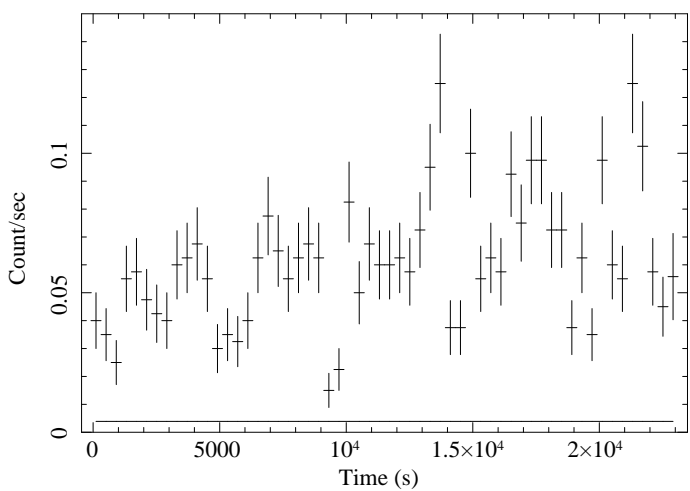}}
 \caption{
    EPIC light curve (0.2-10 keV) of SXP4963 with a binning of 400\,s. The horizontal line at 3.9\expo{-3} \ct\ marks 
    the background level.
  }
  \label{fig:sxp4693_lc}
\end{figure}

SXP4693 - The longest period detected from an SMC BeXRB of 4693\,s was claimed by \citet{2010ApJ...716.1217L} from a 100\,ks 
\cxo\ observation with $\sim$140 net source counts. The pulsar was covered by an observation of the \xmm\ 
survey of the SMC on October 9-10, 2009 \citep[ObsID 0601210801 with a net exposure of about 23\,ks and $\sim$1000 net source counts;][]{2013A&A...558A...3S}.
A Fourier analysis of the light curve reveals a peak at a frequency consistent with the suggested period. We investigated this further using
folding techniques based on $\chi^2$ and Rayleigh Z$^2$ tests \citep{1983A&A...128..245B} 
and a Bayesian periodic signal detection method \citep{1996ApJ...473.1059G} 
as described for instance in \citet{2012MNRAS.424..282C}.
This resulted in a formal pulse period with 1$\sigma$ uncertainty of 4700 $\pm$ 150\,s (see Fig.~\ref{fig:sxp4693_timing}).
The light curve (Fig.~\ref{fig:sxp4693_lc}) shows dips, which repeat every $\sim$4800\,s; they are most likely responsible for the periodicity seen in the timing analysis.
To confirm that this pattern is strictly periodic requires a longer observation, but we conclude that the \xmm\ observation finds evidence
for a period around 4800\,s in agreement with the \cxo\ results presented by \citet{2010ApJ...716.1217L}. 
\citet{2015MNRAS.452..969C} listed the pulsar twice with two different periods as SXP6.62 and SXP4693. We do not see a significant signal 
near 6.62\,s in the \xmm\ data.

\subsubsection{Sources without detected pulsations}

RX\,J0032.9$-$7348 - This hard and variable ROSAT source was discovered by \citet{1996A&A...312..919K} in PSPC observations and proposed as
HMXB candidate. \citet{1999MNRAS.309..421S} found two early-type stars in the ROSAT error circle, one of them showing \Halpha\ emission (their object 1).
Object 2 was identified with GSC0914101338, for which \citep{2004MNRAS.353..601E} give a spectral type of B0.5V.
Object 2 is brighter (B = 15.50 mag, V=15.24) and slightly more distant to the best PSPC position \citep{2000A&A...359..573H}.
Although object 2 cannot be completely ruled out, we assume object 1 as optical counterpart of the X-ray source because of its measured \Halpha\ 
emission line, which suggests a BeXRB nature, but we flag the optical identification as uncertain.

RX\,J0045.6-7313 - \citet{2000A&A...359..573H} suggested this ROSAT source as a BeXRB candidate because an emission-line 
star is located in the X-ray error circle \citep[object 114 in][]{1993A&AS..102..451M}. More recent catalogues of early-type stars in the SMC 
contain additional possible counterpart candidates: \citet{2010AJ....140..416B} performed IR photometry of massive stars and listed AzV9 
\citep{1975A&AS...22..285A} with spectral class B0III \citep[UV spectral classification by][]{1997AJ....114.1951S} 
and IR colours consistent with that of a Be star (see Sect.~\ref{sect:confclass} and Fig.~\ref{fig:SAGE}). 
We add AzV9 to our catalogue and flag the optical identification as uncertain.

CXOU\,J004941.43$-$724843.8 - This source was detected in the Chandra survey of the SMC bar \citep[][ source ID 7\_19]{2009ApJ...697.1695A} 
with the proposed optical counterpart (V = 17.17 mag) 1.36\arcsec\ away from the X-ray position. With an X-ray luminosity of 3.7\ergs{33} 
the source was at low luminosity during the Chandra observation \citep{2014MNRAS.438.2005M}, and it was not detected by \xmm\ \citep{2013A&A...558A...3S}.
\citet{2014MNRAS.438.2005M} presented spectroscopy of
the optical star and reported the detection of a narrow \Halpha\ line, concluding on a spectral type B1-B5 but uncertain Be nature because of the 
marginal line width. All these ambiguities make this case inconclusive.

XMMU\,J005723.4$-$722356 - \citet{2005MNRAS.362..879S} proposed the source detected in \xmm\ data as an HMXB candidate because it might be associated 
with an early-type star. The spectral type of the proposed counterpart was given as B2 (II) by \citet{2004MNRAS.353..601E} and 
confirmed by \citet[][]{2014MNRAS.438.2005M}.  In the SMC catalogue (source 65) of \citet[][]{2013A&A...558A...3S}, the \xmm\ position is 1.73\arcsec\
offset from the optical position given by \citet{2002AJ....123..855Z}. Similarly, the Chandra position is 1.3\arcsec\ away 
\citep{2009ApJ...697.1695A,2014MNRAS.438.2005M}. Both X-ray positions are more consistent with that of an AGN \citep{2013A&A...558A...3S}. 
The source was also detected in three recent \xmm\ observations of SXP5.05 \citep{2015MNRAS.447.2387C} at a much brighter level than in the past. 
During the last observation (ID 0700580601) in particular, with 2.4\ergcm{-13}\ 
the flux was a factor of $\sim$5 higher than the previous maximum during observation 0084200101.
Using the upper limit from observation 0500980201 of 5.8\ergcm{-15} , we derive a variability factor of at least $\sim$40, which
strongly favours an HMXB.
The X-ray position in all three new observations also agrees
better with the B2 star (distances between 0.45\arcsec\ and 1.0\arcsec),
which is further improved when using the target of the observations, SXP5.05 and its optical counterpart, for bore-sight corrections 
(resulting in distances of between 0.36\arcsec\ and 0.68\arcsec).
Optical identification with the early-type star and the strong X-ray variability suggest that XMMU\,J005723.4$-$722356 is an HMXB.

XMM\,J010155.7$-$723236 - Source number 816 in the SMC catalogue of \citet{2013A&A...558A...3S} was originally assigned to SXP7.92. 
However, no pulsations were detected in the \xmm\ data. The improved \xmm\ position is consistent with the Swift source detected 
near AzV285 (see SXP7.92 above) and confirms this star as the
optical counterpart. 
A flux upper limit from one \xmm\ observation with no detection of the source combined with the Swift flux
yields a flux ratio of $>$900, clearly suggesting a BeXRB nature of the source. Analysing OGLE II and III data of AzV285, \citep{2009MNRAS.394.2191C} 
found a 36.79 d period, which was revised by \citet{2011MNRAS.413.1600R} to 36.41$\pm$0.02 d; this is probably the binary period.

XMMU\,J010429.4$-$723136 - Source number 3285 in the SMC catalogue of \citet{2013A&A...558A...3S} was proposed by these authors as an HMXB candidate.
The large X-ray variability and the brightness of the proposed optical counterpart indicates a BeXRB nature of the source.
The source is most likely identical to the Chandra source CXOU\,J010428.7$-$723134 \citep{2011ATel.3154....1R,2013MNRAS.431..252S}, 
although the Chandra position is 2.8\arcsec\ away from that of the proposed counterpart 
\citep[using the coordinates from ][]{2002AJ....123..855Z}. 
The optical star shows a period of 37.15 days in OGLE and MACHO data \citep{2011ATel.3154....1R}, 
which might indicate the orbital period, but could also be an alias of a 0.972 d period, which
\citet{2013MNRAS.431..252S} interpreted as non-radial pulsations of a Be star.
\citet{2011ATel.3154....1R} also reported a period of 707\,s found in the Chandra X-ray data. 
However, this period is most likely an instrumental effect: it is exactly the Chandra dithering period and the source is located at 
the rim of a CCD, moving in and out of the detector. This probably also explains the relatively large angular distance between the Chandra 
and the optical position. 
A likely ROSAT detection ([HFP2000]264) is listed in the PSPC catalogue of \citet{2000A&AS..142...41H}.
We conclude that XMMU\,J010429.4$-$723136 = CXOU\,J010428.7$-$723134 = RX\,J0104.5$-$7231 is a BeXRB in the SMC, 
although measurements of the spin period of the NS and of the strength of the \Halpha\ line are still required for the final confirmation.


\subsubsection{Rejected candidates from previous work}
\label{sect:nocat}

For completeness we provide here a list of former HMXB candidates, which were rejected in earlier work. These objects 
are not included in our catalogue.

XMMU\,J004833.4$-$732355 - Source number 40 (uncertain nature) in Table 2 of \citet{2005MNRAS.362..879S} is classified as an
AGN by \citet{2013A&A...558A...3S}. See source 123 in their catalogue.

XMMU\,J005156.0$-$734151 = RXJ0051.7$-$7341 - Source number 1 of \citet{2003A&A...403..901S} is rejected as an HMXB candidate. 
In the improved error circle of the source derived from additional observations 
\citep[source 100 in the SMC catalogue of][]{2013A&A...558A...3S} no bright optical counterpart with colours of an early-type 
star is found.

XMMU\,J005432.2$-$721809 - Source number 36 (uncertain nature) in Table 2 of \citet{2005MNRAS.362..879S} is newly classified as 
AGN \citep[source 62 in][]{2013A&A...558A...3S}. From the X-ray spectrum a photon index typical for an AGN is derived and the position is
2.7" (4.8$\sigma$) away from the V=16.6 mag star, originally proposed as the counterpart.

XMMU\,J005441.1$-$721720 - Another source of uncertain nature (number 41) from \citet{2005MNRAS.362..879S}. The revised position
\citep[source 82 in][]{2013A&A...558A...3S} is inconsistent with the originally proposed counterpart (V=15.7 mag star), and 
the source is classified as AGN candidate.

CXOU\,J005504.40$-$722230.4 - \citet{2014MNRAS.438.2005M} presented optical spectroscopy of a B star 3.6\arcsec\ away from the Chandra position.
The large angular distance and the faintness (V = 17.86 mag) of this star make a misidentification very likely.
Moreover, from the \xmm\ detection of the X-ray source \citep[source 976 in][]{2013A&A...558A...3S} the most likely counterpart 
is classified as AGN using a mid-IR colour selection \citep{2009ApJ...701..508K}. 

CXOU\,J005527.9$-$721058 was found in Chandra data and suggested as a BeXRB pulsar by \citet{2004ATel..217....1E,2004MNRAS.353.1286E}. 
The source was later detected by \xmm\ (XMMU\,J005527.6$-$721059). 
The two X-ray positions obtained from Chandra and \xmm\, are inconsistent with that of the proposed 
Be star counterpart. Furthermore, the X-ray spectrum is more typical of an AGN, making it highly likely that the period, detected with only 
2.5$\sigma$ confidence, was spurious and the identification with the Be star was incorrect \citep{2008ATel.1529....1H}.

XMMU\,J010016.1$-$720445 = RX\,J0100.2$-$7204 - Additional \xmm\ observations of source number 11 of \citet{2003A&A...403..901S} 
provide an improved position that excludes identification with a bright optical counterpart. The X-ray source is classified as 
an AGN candidate \citep[source 55 in][]{2013A&A...558A...3S}.

XMMU\,J010137.4$-$720418 = RX\,J0101.6$-$7204 - This source is located close to the \xmm\ calibration target 1E\,0102.2$-$7219 and 
was detected more than 30 times. The error-weighted X-ray position is incompatible with that of the emission-line star [MA93]1277
\citep{1993A&AS..102..451M}, which was proposed by \citet{2000A&AS..147...75S} as possible counterpart in a BeXRB. The source
appears as entry 21 in \citet{2005MNRAS.362..879S}. The steepness of the power-law X-ray spectrum and the small X-ray variability 
of a factor of $\sim$4 over 12 years of \xmm\ observations are fully compatible with an AGN nature \citep[source 53 in][]{2013A&A...558A...3S}.

AX\,J0105$-$722 - The detection of a 3.34 s periodicity with 99.5\% confidence (2.8$\sigma$) was claimed by \citet{1998IAUC.7028....1Y}.
The X-ray spectrum was modelled with a power law with photon index 2.2 $\pm$ 0.3, which is much steeper than typically found for HMXBs in the 0.2$-$10 keV band.
RX\,J0105.1$-$7211 was suggested as the most likely ROSAT counterpart of the ASCA source with MA93[1517] \citep{1993A&AS..102..451M} located 7.7\arcsec\ 
away from the ROSAT position \citep{2000A&A...353..129F}. 
Another nearby emission line star (MA93[1506]) was identified as Be star of spectral type B1-B2 III-Ve (with equivalent width of 
the \Halpha\ line of -54\,\ang) and favoured as counterpart
because it has an optical period of 11.09\,d \citep{2005MNRAS.356..502C,2008MNRAS.388.1198M}. However, such a short orbital period is not 
expected from the \Halpha\ - orbital period relation, which suggests on orbital period longer than 100\,d (Fig.\,\ref{fig:porb_Ha}).
\citet{2008A&A...491..841E} used \xmm\ observations to improve 
the X-ray position and classified XMMU\,J010509.7$-$721146 (power-law photon index 2.0 $\pm$ 0.3; no V $<$ 18 optical counterpart in error circle)
as AGN. Positional coincidence and the agreement in X-ray spectral properties strongly suggest that AX\,J0105$-$722, RX\,J0105.1$-$7211 
and XMMU\,J010509.7$-$721146 are the same source and incompatible with [MA93]\,1517 and [MA93]\,1506
as optical counterpart. The 3.34 s periodicity is most 
likely spurious \citep[but still appears in the list of ][]{2015MNRAS.452..969C}. 

XMMU\,J010519.9$-$724943 - As optical counterpart, \citet{2014MNRAS.438.2005M} suggested a B3-B5 star with \Halpha\ emission 
4.0\arcsec\ from the X-ray position determined 
by these authors. Based on astrometric corrections and combining data from two \xmm\ observations, \citet{2013A&A...558A...3S} derived a position 4.9\arcsec\ away from the optical position (source 420 in the \xmm\ catalogue). 
This corresponds to 5$\sigma,$ making the identification highly unlikely. Given the classification as AGN by \citet[][]{2013A&A...558A...3S}, 
we conclude that the Be star is not the counterpart of the X-ray source.

XMMU\,J010620.0$-$724049 - This case is very similar to the previous, here the separation Chandra/optical position is 4.5\arcsec. Again 
\xmm\ detections in two observations provide an X-ray position (source 426 in \xmm\ SMC catalogue) that is even farther away (7.6\arcsec). 
A very likely AGN-counterpart is detected in Spitzer/IRAC images\footnote{SAGE LMC and SMC IRAC Source Catalog (IPAC 2009) available in Vizier} 
located only 0.64\arcsec\ from the \xmm\ position. We conclude that the X-ray source is not a BeXRB, but an AGN.

\section{Confidence classes of HMXBs and candidates}
\label{sect:confclass}

Candidates for HMXBs in the SMC were proposed in various publications mainly based on the positional coincidence of 
an X-ray source with an optical counterpart with appropriate brightness and colours consistent with an early-type star. 
The confidence with which they can be claimed to be a real HMXB or a BeXRB depends in particular on various aspects, such as  the error of the X-ray position 
(usually significantly larger than the uncertainty in the optical position) and properties of the X-ray source as well as the 
candidate optical counterpart. This includes spectral and timing properties of the X-ray source and the optical star. 
When several criteria can be applied successfully, the confidence increases for a correct association of the X-ray source 
and the optical counterpart and its identification as an HMXB. Following this, we assigned flags to the 
(candidate) HMXBs in our compiled catalogue as described in Table~\ref{tab:flags}.

Based on the flags, we divided the list of HMXBs and candidates into six confidence classes, which are summarised in Table~\ref{tab:classes}. 
Sources with detected pulse period in X-rays (class I with flag ps or ps:) are most likely HMXBs even when the optical counterpart is not clearly 
identified as yet because the uncertainties in the X-ray position
are large. X-ray sources with large long-term variability 
or a hard spectrum as typical for BeXRBs \citep[in the 0.2-10 keV band the photon index is lower than $\sim$1.3;][]{2008A&A...489..327H} 
can also be identified as HMXBs from their X-ray properties (class II with flag xv or xs). 
A third class with secure identification of the optical 
counterpart (due to a precise X-ray position) as an early-type emission-line star (flags oi and em) are also BeXRBs with 
high confidence. With increasing uncertainty in the X-ray position, the chance coincidence for the presence of an early-type 
star in the error circle increases. A larger X-ray position error is typically found for weak X-ray sources, which do 
not allow deriving other X-ray properties either. In class IV we list X-ray sources with less certain position and an 
emission-line star in the error circle (flags oi: and em). Class V comprises X-ray sources with a good position 
and therefore high confidence for their identification with early-type stars (flag oi). 
However, no information about \Halpha\ emission is available. 
Only three such sources are found in our catalogue, and optical spectra should easily allow confirming their probable Be-star counterpart.
Finally, in class VI the highest fraction of chance correlations of the X-ray sources with early-type stars is 
expected. These are usually weak X-ray sources (flag oi: with position errors typically larger than 1$\arcsec$), and apart from the 
appropriate colours, no other information about the possible optical counterpart is available.

\begin{table}
\caption[]{Flags indicating different properties as measured for HMXBs and candidates.}
\begin{center}
\begin{tabular}{lll}
\hline\hline\noalign{\smallskip}
\multicolumn{1}{l}{C\tablefootmark{a}} &
\multicolumn{1}{l}{Flag\tablefootmark{b}} &
\multicolumn{1}{l}{Description}\\
\noalign{\smallskip}\hline\noalign{\smallskip}
 8 & ps  & X-ray modulation indicates NS spin period \\
 9 & px  & long-term X-ray period suggests orbital period \\
10 & po  & period in optical suggests orbital period \\
11 & os  & orbital solution available\\
11 & ox  & assuming X-ray period as orbital period\\
11 & oo  & assuming optical period as orbital period\\
11 & oxo & X-ray and optical period are consistent\\
12 & xv  & variability in X-rays larger than a factor 30 \\
13 & xs  & typical X-ray spectrum\\
   &     & (power law with photon index $<$ 1.3)\\
14 & oi  & optical id with high confidence \\
15 & em  & Balmer (\Halpha) emission measured from spectrum\\
15 & nem & no near-IR excess emission \\
\noalign{\smallskip}\hline
\end{tabular}
\tablefoot{
\tablefoottext{a}{Column in catalogue table (see Table~\ref{tab:catdesc}).}
\tablefoottext{b}{A colon behind the flag indicates an uncertain property.}
}
\end{center}
\label{tab:flags}
\end{table}

\begin{table}
\caption[]{Summary of confidence classes.}
\begin{center}
\begin{tabular}{lll}
\hline\hline\noalign{\smallskip}
\multicolumn{1}{l}{Class} &
\multicolumn{1}{l}{Flags} &
\multicolumn{1}{l}{Number of sources}\\
\noalign{\smallskip}\hline\noalign{\smallskip}
I   & ps $\cup$ ps:    & 63 \\
II  & xv $\cup$ xs     & 18 \\
III & oi $\cap$ em     & 12 \\
IV  & oi: $\cap$ em    &  8 \\
V   & oi               &  3 \\
VI  & oi:              & 44 \\
\noalign{\smallskip}\hline
\end{tabular}
\end{center}
\label{tab:classes}
\end{table}

As mentioned above, the probability for false identification of optical counterparts to the X-ray sources in class VI is higher than for classes I-V.
The optical counterparts of the large majority of the BeXRB pulsars in class I are well investigated. Spectral types and the strength of the \Halpha\
emission were determined and can be found in the literature \citep[e.g.][]{2005MNRAS.356..502C,2008MNRAS.388.1198M}. Therefore, class I objects can 
be used to define a representative sample. Comparison of various source parameters between the different classes may then reveal differences, which 
indicate false identifications.

\begin{figure}
  \resizebox{\hsize}{!}{\includegraphics[clip=]{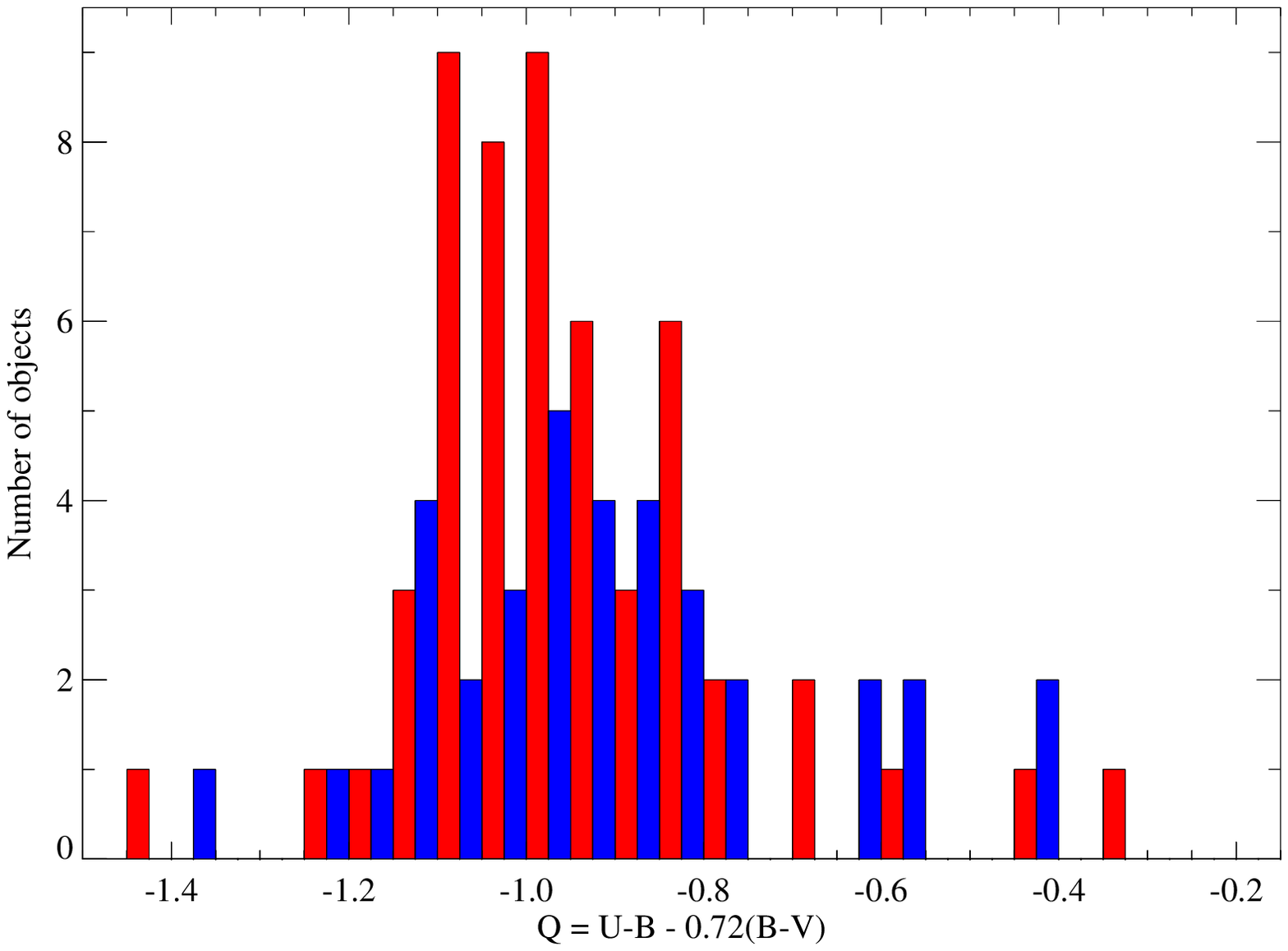}}
  \resizebox{\hsize}{!}{\includegraphics[clip=]{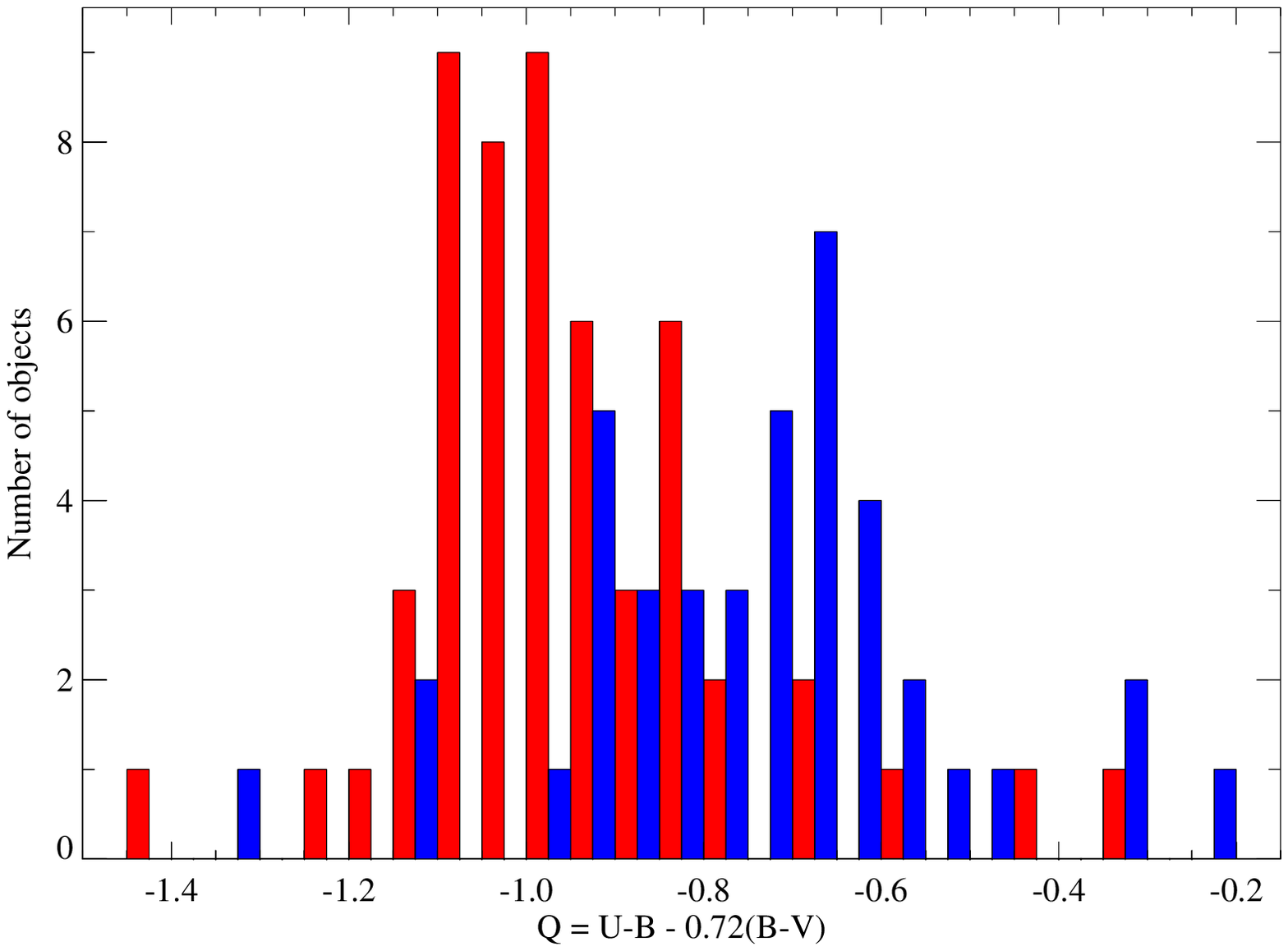}}
  \caption{
    Distribution of the reddening-free Q parameter for different HMXB confidence classes. 
    The top panel compares the pulsars (class I) with identified optical counterpart (red) with objects from classes II to V (blue), 
    while in the bottom panel the comparison with class VI is shown.
  }
  \label{fig:hist_qpar}
\end{figure}

When we plotted colour-colour (U-B vs. B-V) and colour-magnitude (V vs. B-V) diagrams, we found evidence that the objects in class VI are optically 
fainter and redder.
To make this quantitatively clearer, we used the reddening-free Q-factor Q=U-B-0.72(B-V) and compared its distribution between the different classes. 
While classes II-V are statistically consistent with class I, class VI exhibits a significantly different distribution towards higher Q values. 
This is demonstrated in Fig.~\ref{fig:hist_qpar} where the distribution of the Q parameter is compared for class I and class II-V (top panel) and 
for class I and class VI (bottom). A statistical Kolmogorov-Smirnov test results in a probability of 98.5\% for class I and II-V  and 
1.3\expo{-7} for class I and VI to be drawn from the same distribution.
This clearly shows that the early-type stars proposed as optical counterparts of the X-ray sources from class VI are on average of different (later) 
spectral type than the optical counterparts of class I, which show a relatively narrow distribution around B0-B1 \citep[][]{2008MNRAS.388.1198M}. 
In particular, no spectral type later than B2 was found for BeXRBs with well-known optical counterpart in the Milky Way \citep[][]{2011Ap&SS.332....1R}.
On the other hand, isolated Be stars with later spectral type exist \citep[][]{2008MNRAS.388.1198M}, and we cannot exclude that (in particular X-ray faint)
class VI objects might be associated with a late-type BeXRB.
However, the larger X-ray error circles of the faint class VI objects implies that many of them are chance coincidences of the X-ray sources with 
single late B- or even Be-type stars.

\begin{figure*}
  \resizebox{\hsize}{!}{\includegraphics[clip=]{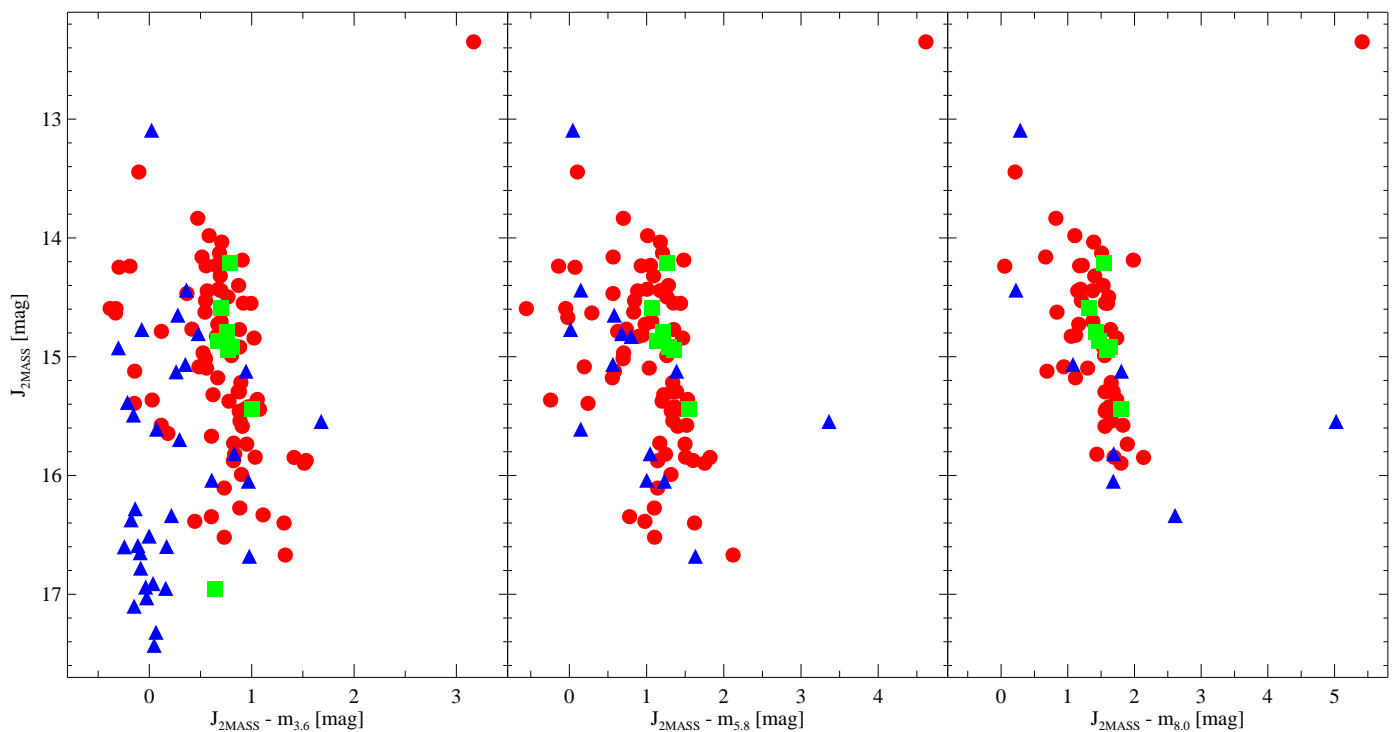}}
  \caption{
    Infrared excess using the three IRAC magnitudes at 3.6, 5.8, and 8.0 $\mu$m and J band from 2MASS (in a few cases from IRSF when no 2MASS value was available).
    Blue triangles mark sources from class VI (flag ``oi:''), green squares those from class IV (with additional flag ``em''), while red circles indicate flag ``oi''.  
  }
  \label{fig:SAGE}
\end{figure*}

\citet{2010AJ....140..416B} have used Spitzer IRAC fluxes (at 3.6, 4.5, 5.8, and 8.0 $\mu$m) of O and early-B stars in the SMC to distinguish 
OBe from OB stars based on their infrared excess relative to J-band fluxes. In Fig.~\ref{fig:SAGE} we plot the infrared excess for the catalogue sources
with flags ``oi'' and ``oi:''. A remarkably large portion of sources from class VI (16 out of 44) fall in a region with J$>$16 and J-m$_{3.6}<0.3$ 
where no other sources are found. This suggests that these 16 objects are most likely normal B stars without a circum-stellar disc 
\citep[][where a threshold of J-m$_{3.6}>0.5$ is defined to introduce a photometric Be star classification]{2010AJ....140..416B}. 
Fourteen of the sixteen objects are from the BeXRB candidate list derived from the XMM-Newton survey of the SMC 
\citep[sources 41, 259, 468, 474, 1019, 1189, 1762, 1955, 2211, 2569, 2675, 2721, and 2967, 3271 from][]{2013A&A...558A...3S}. 
The authors determined the number of chance coincidences as high as twenty. Therefore, the lack of infrared excess excludes the
greater part of the expected chance coincidences.
The remaining two objects (CXO\,J005331.8-721845 and CXO\,J005419.2-722049) are from the candidate list of \citet{2010ApJ...716.1217L}, very faint X-ray 
sources with positional errors larger than 5\arcsec. Neither are found in the CSC. The proposed optical counterparts have optical colours consistent 
with an early-type star. 
They do not show infrared excess, which suggests that they are not Be stars and therefore are most unlikely the counterparts of the weak 
Chandra sources, which might even be spurious detections after all.
This is also supported by their Q-values, which are $>-0.91$.
We flag these sixteen objects with ``nem'' in the catalogue. 

As expected, the class IV objects are among the well-established BeXRBs, as is shown by their flag ``em''. However, it should be noted that for class IV objects 
a misidentification of the X-ray source with a Be star can still not be excluded: As a result of their larger X-ray position uncertainty, the number of chance 
coincidences is higher than for objects whose optical counterpart is unambiguously identified (flag ``oi'').

\section{Candidates that are probably misidentified as HMXBs}
\label{sect:rejected}

In Sect.~\ref{sect:confclass} we have shown that a large portion of the 44 sources in confidence class VI is 
most likely misidentified as HMXBs. 
New data for some of the proposed candidates further contradict their proposed HMXB nature. In the following we list objects
that we finally consider to probably not be HMXBs or which we reject.

\citet{2013A&A...558A...3S} present a list of 45 HMXB candidates selected from their SMC X-ray source catalogue (see their Table\,5).
Taking into account the positional uncertainties of the X-ray sources, they estimate 16.6 $\pm$ 3.4 chance coincidences.
Meanwhile, additional information is available for many of the candidates, which allows us to better constrain the origin of their X-ray emission.
This includes Chandra X-ray data \citep[with source positions often available from the on-line catalogue of][]{2010ApJS..189...37E}, possible 
AGN counterparts properly selected from Spitzer data using colour indices (SAGE LMC and SMC IRAC Source Catalog - IPAC 2009), and absence 
of a near-IR excess, normally seen from Be stars (see Sect.~\ref{sect:confclass}).
In particular, \citet{McBride2016} have obtained optical spectroscopy of all candidates presented by \citet{2013A&A...558A...3S}, and in 
several cases no Balmer emission was found (although \Halpha\ was not always covered). 
It should be noted here that Be stars can lose and rebuild their circum-stellar disc, as is indicated by the disc emission coming 
and going on timescales of decades \citep{2013A&ARv..21...69R} 
with state transitions as short as a few months \citep[e.g.][]{2007A&A...462.1081R}. 
Although we consider a disc loss between X-ray and optical observations unlikely, we cannot exclude that a BeXRB is observed during 
a disc-less state. 
Therefore, we label cases without an indication of a circum-stellar disc because there is no near-IR excess {\it or} Balmer emission 
as ``HMXB unlikely'' and only cases without near-IR excess {\it and} no Balmer emission as ``HMXB rejected'' in the comment 
column of the catalogue. We kept the unlikely/rejected cases in our catalogue.
We also rejected candidates with improved X-ray positions that exclude the previously suggested counterparts, candidates that
are more probably AGN because of their Spitzer counterparts and other information (summarised below).

At the current status of our work, this means that we have 27 candidates in total 
(11 marked as unlikely and 16 as rejected) that probably are HMXB misidentifications. 
Twenty-five of them belong to confidence class VI, only one was given 
class III ([SHP2013]\,1408, see below) and one class V ([SHP2013]\,287, see below). 
Twenty of the twenty-seven sources are from the candidate list of \citet{2013A&A...558A...3S}.
From this list we can confirm fourteen sources as HMXBs, and for eleven cases no additional 
information is available and they remain candidates.
Following SIMBAD naming conventions, we use [SHP2013]\,N for source number N in the catalogue of 
\citet{2013A&A...558A...3S} and [SG2005]\,SMC\,N for sources in Table\,2 of \citet{2005MNRAS.362..879S}. 

For [SHP2013]\,117 = [SG2005]\,SMC\,34 a V=16.2 mag star was proposed as the optical counterpart. \citet{2013A&A...558A.101S} argued 
for QSO J004818.72-732059.83 \citep{2009ApJ...701..508K} as another likely counterpart, very close to the X-ray position. 
\citet{McBride2016} 
found no Balmer emission from the early-type star, making the QSO the more likely counterpart.

A V=16.7 mag star \citep[with spectral type B2\,V][]{2003A&A...401L..13N} was proposed as counterpart for [SHP2013]\,160 = [SG2005]\,SMC\,39.
No Balmer emission \cite{McBride2016} and the presence of a Spitzer counterpart make an AGN nature more likely.

[SHP2013]\,259 = [SG2005]\,SMC\,48 was identified as an eclipsing binary with an orbital period of 5.18 days \citep{2004AcA....54....1W}. This 
makes an HMXB nature highly unlikely.

A V=15.1 mag star \citep[with spectral type B1-3\,III from the 2dF survey of the SMC,][]{2004MNRAS.353..601E} was proposed as the counterpart 
for [SHP2013]\,287. \citet{McBride2016} 
found no Balmer emission from the early-type star, suggesting a chance coincidence with the X-ray source.

The proposed counterpart of [SHP2013]\,474 shows no near-IR excess (see Sect.~\ref{sect:confclass}), and \citet{McBride2016} 
found no Balmer emission.

For [SHP2013]\,562 a V=15.4 mag star with spectral type B0\,V from the 2dF survey \citep{2004MNRAS.353..601E} was proposed as counterpart. 
\citet{McBride2016} 
found no Balmer emission. 

The angular distance between [SHP2013]\,1019 and its proposed optical counterpart is 3.5\arcsec. 
\citet{McBride2016} 
found no Balmer emission. 
No near-IR excess is seen either (see Sect.~\ref{sect:confclass}). 
The presence of a Spitzer counterpart makes a chance coincidence with the early-type star most likely and suggests an AGN origin for 
the X-ray source.

[SHP2013]\,1189 is located about 3.5\arcmin\ from 1E0102.2-7219, the X-ray brightest supernova remnant in the SMC that is frequently observed
by \xmm. The faint source was detected only once with a large positional error, and no entry is found in the Chandra on-line catalogue 
near its position. However, the source is clearly seen in the merged Chandra ACIS-I image presented in \citet{2013A&A...558A...3S},
which suggests a constantly faint source. 
The Chandra position is incompatible with that of the V=16.1 mag star proposed as counterpart. This star also shows no near-IR excess.

[SHP2013]\,1408 is associated with the B[e] super-giant star LHA\,115-S\,18 \citep{2013A&A...560A..10C,2014MNRAS.438.2005M}, and the X-ray 
emission is probably not produced by accretion onto a compact object.

The case of [SHP2013]\,1762 is very similar to [SHP2013]\,1189: near 1E0102.2-7219; seen in the merged Chandra ACIS-I image; 
Chandra position incompatible with the V=16.3 mag star proposed as counterpart, which shows no near-IR excess.

\citet{McBride2016} identified the proposed counterpart to [SHP2013]\,1823 as Be star.
However, the 
uncertain \xmm\ position is at a distance of 4.9\arcsec. The source was also detected in Chandra data \citep{2010ApJ...716.1217L}, also 
5\arcsec\ from the optical position. The Chandra 95\% uncertainty of 1.83\arcsec\ makes the identification of the X-ray source with 
the Be highly unlikely. Moreover, the existence of a Spitzer counterpart suggests an AGN origin.

The \xmm\ detection of the weak source [SHP2013]\,1826 yields a highly uncertain position and is 5.7\arcsec\ away from 
its proposed optical counterpart. In addition, a Chandra detection \citep[95\% error = 0.47\arcsec, distance 5.4\arcsec\ ][]{2010ApJ...716.1217L}
excludes an identification with the proposed V=14.9 mag star.

The proposed counterpart of [SHP2013]\,1955 shows no near-IR excess (Sect.~\ref{sect:confclass}), and \citet{McBride2016} 
found no Balmer emission. We conclude that the X-ray source is not associated with the normal B star.

The angular distance between [SHP2013]\,2100 and its proposed optical counterpart is 4.3\arcsec\ , and \citet{McBride2016}
found no Balmer emission.

The position of the proposed counterpart of [SHP2013]\,2318 is 5.7\arcsec\ away. We determined an improved position from a recent 
deep Chandra observation (ID 14671) to an accuracy of 0.3\arcsec\ (1 $\sigma$). The Chandra position is within 2.0\arcsec\ of the \xmm\ 
position, but with 6.9\arcsec\ distance incompatible with that of the V=16.1 mag early-type star.

\citet{McBride2016} found no Balmer emission from the suggested counterpart of [SHP2013]\,2497, which makes a chance 
correlation with a normal B star most likely.

No near-IR excess and no Balmer emission are seen from the suggested counterpart of [SHP2013]\,2569 \citep{McBride2016}.
The existence of a Spitzer counterpart suggests an AGN origin.

[SHP2013]\,2675 is another case that was identified as an eclipsing binary \citep[orbital period of 3.29 days][]{2004AcA....54....1W}, 
which makes an HMXB nature highly unlikely. No near-IR excess is seen from the binary star system.

For [SHP2013]\,2737 a V=14.7 mag star with spectral type B5\,II from the 2dF survey \citep{2004MNRAS.353..601E} was proposed as counterpart. 
\citet{McBride2016} 
found no Balmer emission. A Chandra detection \citep{2010ApJS..189...37E} 
yields a position 2.9\arcsec\ from the B star. A Spitzer counterpart suggests an AGN origin.

The V=16.6 mag star suggested as counterpart for [SHP2013]\,3271 is 4.4\arcsec\ away from the \xmm\ position. 
\citet{McBride2016} 
found no \Halpha\ emission. No detected near-IR excess is 
consistent with a normal B star found by chance in the X-ray error circle.

\section{Population statistics}
\label{sec:population}

\begin{figure}
  \resizebox{\hsize}{!}{\includegraphics[clip=]{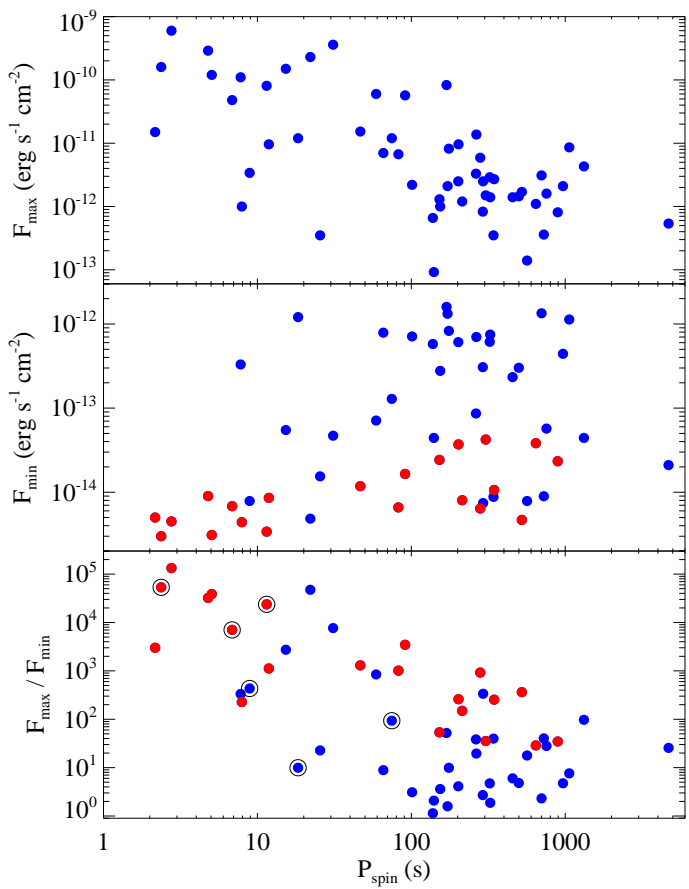}}
  \caption{
    Observed maximum (top) and minimum (middle) X-ray flux together with their ratio (bottom) as function of spin period. 
    When sources were not detected in a second observation, available upper limits were used for the minimum flux (marked in red).
    Black circles in the bottom panel indicate the BeXRBs with measured orbit eccentricity.
  }
  \label{fig:fmax_fmin}
\end{figure}

\begin{figure}
  \resizebox{\hsize}{!}{\includegraphics[clip=]{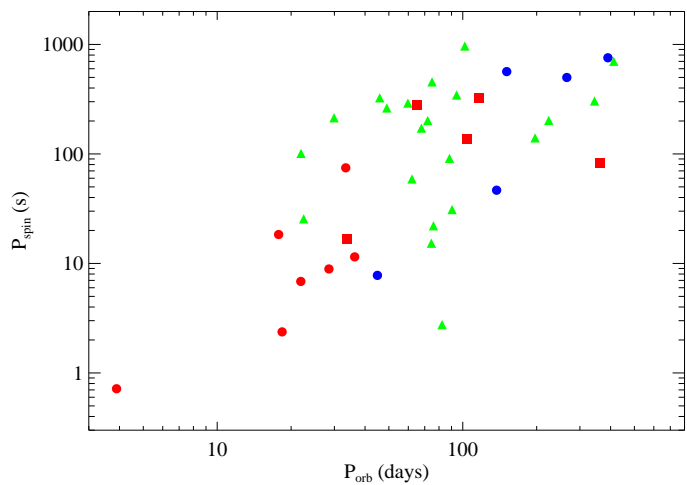}}
  \caption{
    Spin period vs. orbital period for HMXBs in the SMC. The super-giant HMXB SMC\,X-1 is found in the lower-left corner.
    Orbital periods found with different methods are marked with different symbols: orbit solution (red circles), X-ray outbursts (red squares), 
    optical light curve (green triangles). Blue circles mark systems with orbital periods consistently derived from X-rays and optical.
}
  \label{fig:porb_spin}
\end{figure}

The large number of HMXBs in the SMC allows statistical investigations of their X-ray and optical properties. Many BeXRBs were found because coherent pulsations in the X-ray flux (often during outburst)
were detected, which indicates the spin period of the NS. 
\citet{2011Natur.479..372K} discussed the bimodal distribution of the spin period with two maxima at around 10\,s and between 100\,s to 1000\,s 
in terms of two populations of X-ray pulsars produced by two types of supernovae. 
As an alternative explanation, \citet{2014ApJ...786..128C} proposed that the spin period distribution is the result of two different accretion modes.
To elaborate on this question, we collected maximum and minimum fluxes (or upper limits when lower) reported for the SMC BeXRBs in the 
literature and entered them in our catalogue. 
The high sensitivity of the \xmm\ SMC survey provides stringent upper limits for non-detected sources. These are available in the 
catalogue of \citet{2013A&A...558A...3S} for sources observed more than once and detected at least once. For the remaining SMC HMXBs, which 
were never detected in any \xmm\ observation, we readout the upper limits from the sensitivity maps produced for the work of 
\citet{2013A&A...558A...3S}. If an \xmm\ upper limit was higher than any detected (minimum) flux from another instrument, then 
it was not used. 
In Fig.~\ref{fig:fmax_fmin} we plot the maximum and minimum fluxes ($\fmax$ and $\fmin$) and the derived ratio 
(variability factor) for the SMC BeXRB pulsars versus 
their spin period. As shown by \citet{2014ApJ...786..128C}, short-period ($<$40 s) pulsars show on average higher maximum luminosities than 
long-period pulsars. This is demonstrated by the rather loose anti-correlation of maximum flux with spin period in the upper panel of 
Fig.~\ref{fig:fmax_fmin}. We note that the minimum observed flux (or upper limit for non-detections) also tends to be lower for the 
short-period pulsars. While most of the short-period pulsars even fall below the detection limits of modern X-ray instrumentation, many 
long-period pulsars are always detected well above these limits (Fig.~\ref{fig:fmax_fmin}, middle panel). As a consequence, 
the anti-correlation of $\fmax$/$\fmin$ with spin period is even more pronounced (Fig.~\ref{fig:fmax_fmin}, lower panel). 
There still might be some observational bias in the correlations, since some systems have not been caught at outburst maximum or at 
their minimum flux level and the observed $\fmax$/$\fmin$ value presents only a lower limit. 
However, given the large number of X-ray observations, this should affect only a relatively small number of 
objects independent of spin period and would not change the overall distributions. 
It is therefore remarkable that the upper envelope of the data points in the bottom panel of Fig.~\ref{fig:fmax_fmin} is so sharp, 
suggesting that the neutron stars with longer spin periods - and wider orbits (Fig.~\ref{fig:porb_spin}) - sample a narrower range in accretion rate than 
short-period pulsars. The largest variations in accretion rate are expected for systems with high eccentricity and/or large tilt between
orbital plane and Be disc.

For six of SMC XRBs the eccentricity of the orbit is known with values between 0.07 and 0.43. Interestingly, among these, SMC\,X-2 has the 
lowest eccentricity and the highest variability factor of $\sim$5\expo{4}. Although the orbital parameters could be determined only for neutron stars with short spin periods
(which show the highest variability), the measured eccentricities are relatively moderate. 
This may indicate that eccentricity is not the main parameter leading to the strong outbursts from these systems. Alternatively, a considerable inclination of the orbit with respect to the plane of the circum-stellar disk can also lead to strong and in particular short
outbursts when the NS passes through the disc. 
Strong evidence for this kind of system geometry was found for SXP5.05 \citep{2015MNRAS.447.2387C}. 
A high inclination with the NS moving far above and below the disc can also explain the lower minimum X-ray flux values observed from 
systems with short spin period (and high variability). 
Moreover, such a system geometry is thought to cause a strong warping of the disc, which then can lead to the strong X-ray outbursts 
and high spin-up of the NS to short periods \citep{2014ApJ...786..128C}. 

While maximum fluxes seen from BeXRBs are mainly determined by the available supply of matter along the neutron star orbit, 
the observed minimum flux is very likely influenced by another mechanism. The propeller effect can inhibit accretion
when the matter from the accretion disc couples onto the rotating magnetosphere of the neutron star at distances larger 
than the co-rotation radius \citep{1986ApJ...308..669S}. The critical luminosity strongly depends on the spin period (Eq.\,5
of that paper) and is expected to be about 2.7\ergs{36} for instance
for SMC\,X-2 with a period of 2.37\,s. 
The large variability factors of short-period pulsars might be explained by this mechanism. When magnetospheric accretion stops,
only faint emission from the (cooling) neutron star surface remains, possibly with some small contribution from direct accretion.
Several BeXRB pulsars observed in extreme low states at X-ray luminosities of $\sim$\oergs{34} and 
below are believed to be seen in the propeller state \citep[see][and references therein]{2005A&AT...24..151R}.
For long-period pulsars the propeller effect is thought to play no role (e.g. for a period of 400\,s the critical luminosity is at 
1.7\ergs{31}).  However, we note here that \citet{2008A&A...491..841E} reported a sharp drop in X-ray luminosity below the detection 
limit of \xmm\ from SAX\,J0103.2-7209. The drop occurred after the source reached a level of constant spin period (after years of spin-up) 
at a luminosity of about 3\ergs{35} and a small further decrease to 2\ergs{35}. The authors suggest that this behaviour could be 
qualitatively explained by the propeller effect, but at a critical luminosity four orders of magnitude higher than expected from 
Eq.\,5 of \citet{1986ApJ...308..669S}.

\begin{figure}
  \resizebox{\hsize}{!}{\includegraphics[clip=]{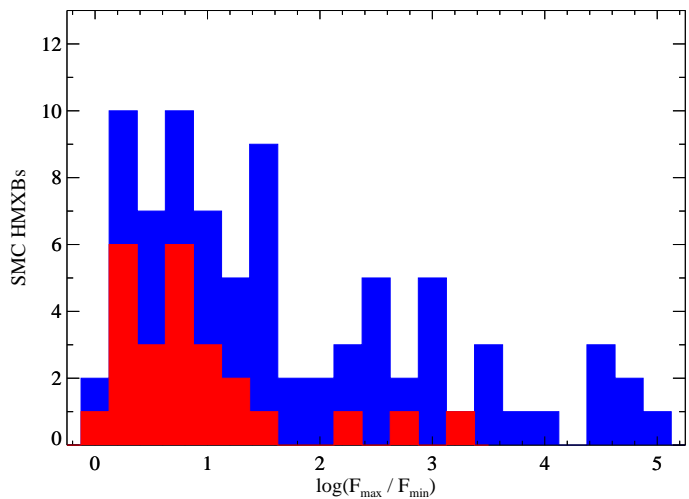}}
  \caption{
    Distribution of the variability factor $\fmax$/$\fmin$ for all HMXBs with confidence class I-V (blue) and for the 
    non-pulsars (confidence class II-V, red) in the SMC.
  }
  \label{fig:hist_var}
\end{figure}

\begin{figure}
  \resizebox{\hsize}{!}{\includegraphics[clip=]{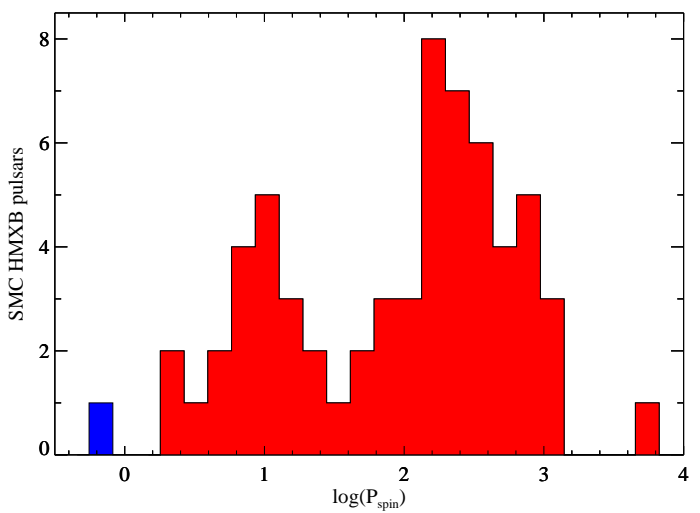}}
  \caption{
    Spin period histogram for 63 HMXB pulsars in the SMC. BeXRBs are shown in red, while the super-giant system SMC\,X-1 is depicted in blue.
  }
  \label{fig:hist_spin}
\end{figure}

Figure~\ref{fig:hist_var} demonstrates that the distributions of the variability factor among BeXRB pulsars and sources without detected 
pulsations in confidence classes II-V are also similar. If the anti-correlation of variability and spin period also holds for 
the sources that have no detected spin period (yet), this suggests that most class II-V sources probably have longer spin 
periods, which would contribute to the long-period peak in the spin-period distribution of Fig.~\ref{fig:hist_spin}.
Because sufficiently long observations are required to detect these periods, this could be one reason why they have not yet
been found.
Conversely, only three (one) class II-V objects have variability factors greater than 100 (1000). For all three, few X-ray observations 
exist in general, and none have sufficient statistics to allow a sensitive period search.

In Fig.~\ref{fig:hist_spin} we present the updated spin period histogram for SMC pulsars, 
which also includes the two objects with low-significance detections of the period, which need to be confirmed: 
the 6.878\,s detected in Integral data \citep{2007MNRAS.382..743M} and 
the 154\,s period from XMMU\,J010743.1$-$715953 \citep{2012MNRAS.424..282C}.
We also include the long 4693\,s period found by \citet{2010ApJ...716.1217L} from a 100\,ks \cxo\ observation that may be present 
in a 23\,ks \xmm\ observation (see Sect. 2.2).

Orbital periods of HMXBs and in particular from BeXRBs can be identified from their long-term light curves in X-rays and optical.
Long-term monitoring of the SMC pulsars in X-rays was performed by RXTE \citep{2008ApJS..177..189G}, which was sensitive to periodic 
outbursts near periastron passage of the NS. The OGLE project has monitored stars in the SMC for more than a decade and 
revealed periodic long-term variability for many Be stars in BeXRBs, which can be attributed to the orbital period of the 
binary system \citep{2011MNRAS.413.1600R}. For a few systems the Doppler analysis of the X-ray pulse timing data allowed deriving 
orbital solutions \citep[e.g.][]{2011MNRAS.416.1556T}. 
In Fig.~\ref{fig:porb_spin} we show the spin versus orbital period diagram for the HMXBs in the SMC methodised by different symbols.
The first version of such a diagram \citep{1984A&A...141...91C} showed a strong correlation between the two parameters for BeXRBs.
Our most recent version contains five times more systems and the correlation has considerably weakened.
The correlation is thought to be explained by a quasi-equilibrium state of the neutron star in which the co-rotation radius 
equals the Alfv{\'e}n radius. In this picture, the rotation period and the magnetic field strength of the neutron star and the 
accretion rate determine whether matter can be accreted onto the neutron star (and spin it up) or not (leading to spin-down).
The large scatter seen in our updated diagram (the spread in spin period for orbital periods between 
50 and 100 days is $\sim$2.5 orders of magnitude) suggests that many of the neutron stars do not rotate near their equilibrium rate, 
even bearing in mind that the model is too simple.

\begin{figure}
  \resizebox{\hsize}{!}{\includegraphics[clip=]{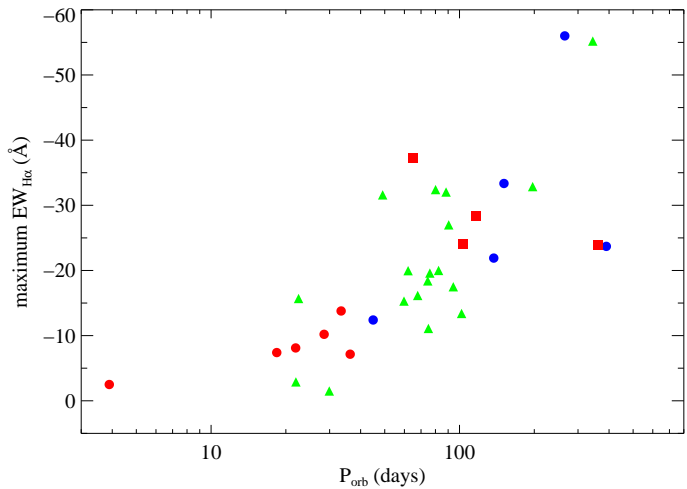}}
  \caption{
    Equivalent width of the \Halpha\ line vs. orbital period for HMXBs in the SMC. The symbols mark the different origin of orbital periods as in Fig.~\ref{fig:porb_spin}.
    When more than one measurement of the equivalent width is available the largest (most negative) value is used.
}
  \label{fig:porb_Ha}
\end{figure}

In Fig.~\ref{fig:porb_Ha} we present another diagram that is
important for characterising the properties of BeXRBs; 
the equivalent width (observed maximum) is plotted as function of orbital period. For all except one of the 
objects in this figure, that is, all whose orbital period is known and that have \Halpha\ measurements, the spin period is known as well.
The exception is the unclear case of RX\,J0049.2-7311, which we do not associate with SXP9.13 (see Sect.\ref{sect:pulsar_notes}).
If the minimum equivalent width is used for objects with more than one \Halpha\ measurement, the plot is not significantly
altered.
\citet{2015MNRAS.452..969C} investigated the dependence of the
equivalent width on orbital period in more detail, in particular with 
respect to the size of the circum-stellar disc and disc truncation by the compact object \citep[see also][]{1997A&A...322..193R}. 
Since in their work SXP9.13 is identified with RX\,J0049.2-7311, our whole sample from Fig.~\ref{fig:porb_Ha} is covered in 
their analysis, and we refer to that work for more details.


\section{Conclusions}
\label{sec:conclusion}

We investigated the properties of 148 (candidate) HMXBs in the SMC and catalogued them.
We assigned different levels of confidence at which they are genuine HMXB systems for all catalogue entries. Pulsars are confidence class I, and the probability for being HMXBs decreases until class VI objects.
We found many chance coincidences between X-ray and optical position in class VI 
and rejected 27 candidates as probably misidentified with normal B stars.

The remaining 121 sources comprise a relatively clean sample of HMXBs in the SMC. With 63 pulsars,
this indicates that almost 50\% of the sources do not have a detected pulse period. A comparison of X-ray 
variability as function of spin period for pulsars and ``non-pulsars'' suggests that many long spin 
periods (longer than a few 100\,s) have probably not been found as yet because of the intrinsic short-term X-ray 
variability of BeXRBs and insufficient observation time. Therefore, it remains unclear how many neutron 
stars in BeXRBs do not exhibit pulsations because their magnetic field axis is (nearly) aligned with the 
rotation axis.
The larger long-term X-ray variability
of objects with short spin period also indicates different accretion schemes for short and long 
spin period pulsars, which may favour the accretion model of \citet{2014ApJ...786..128C} to explain the 
bimodal distribution of spin periods observed from BeXRBs.

\begin{acknowledgements}
RS acknowledges support from the BMWI/DLR grant FKZ 50 OR 0907 during his stay at MPE.
\end{acknowledgements}

\bibliographystyle{aa}
\bibliography{../../../../bibtex/general}

\Online
\scriptsize

\onecolumn
\begin{longtab}
\begin{landscape}
\begin{longtable}{lrrrrrrrrrrrrrp{7cm}l}
\caption{\label{tab:smccat}Catalogue of HMXBs and candidates in the SMC. 
         The full version of the catalogue is available at the CDS and regular updates will be posted at http://www.mpe.mpg.de/heg/SMC}\\
\hline\hline\noalign{\smallskip}
\multicolumn{1}{l}{No} &
\multicolumn{1}{l}{RA} &
\multicolumn{1}{l}{DEC} &
\multicolumn{1}{l}{ERR} &
\multicolumn{1}{l}{O} &
\multicolumn{1}{l}{MA93} &
\multicolumn{1}{l}{U} &
\multicolumn{1}{l}{B} &
\multicolumn{1}{l}{V} &
\multicolumn{1}{l}{I} &
\multicolumn{1}{l}{Q} &
\multicolumn{1}{l}{eQ} &
\multicolumn{1}{l}{D} &
\multicolumn{1}{l}{Pspin} &
\multicolumn{1}{l}{ID} &
\multicolumn{1}{l}{conf. class and flags} \\
\noalign{\smallskip}\hline\hline\noalign{\smallskip}
\endfirsthead
 \hline\hline\noalign{\smallskip}
\multicolumn{1}{l}{No} &
\multicolumn{1}{l}{RA} &
\multicolumn{1}{l}{DEC} &
\multicolumn{1}{l}{ERR} &
\multicolumn{1}{l}{O} &
\multicolumn{1}{l}{MA93} &
\multicolumn{1}{l}{U} &
\multicolumn{1}{l}{B} &
\multicolumn{1}{l}{V} &
\multicolumn{1}{l}{I} &
\multicolumn{1}{l}{Q} &
\multicolumn{1}{l}{eQ} &
\multicolumn{1}{l}{D} &
\multicolumn{1}{l}{Pspin} &
\multicolumn{1}{l}{ID} &
\multicolumn{1}{l}{conf. class and flags} \\
\noalign{\smallskip}\hline\hline\noalign{\smallskip}
\endhead
  1& 01 17 05.2 & -73 26 36 &  0.5 & N &      &         12.23 & 13.01 & 13.15 & 13.17  &  -0.67 &  0.16 &   0.8 &    0.717 & HMXB supergiant SMC X-1                                                          & 1 ps                  oi  em  os  \\ 
  2& 01 19 00.0 & -73 12 27 &220.0 & X & 1868 &         14.27 & 14.96 & 15.04 &        &        &       & 222.7 &    2.165 & HMXB Be/X? XTEJ0119-731 (GCC08), counterpart? Lin 526, opt: M02                  & 1 ps                  oi: em      \\ 
  3& 00 54 30.9 & -73 40 55 & 10.5 & R &      &         13.77 & 14.75 & 14.73 & 14.67  &  -0.99 &  0.06 &  12.4 &    2.370 & HMXB Be/X SMC X-2 (no XMM); likely optical counterp. from OGLE-III variability   & 1 ps  px  po  xv      oi  em  os  \\ 
  4& 00 59 11.4 & -71 38 45 &  7.5 & R & -179 &         13.09 & 14.07 & 14.01 & 14.03  &  -1.02 &  0.08 &   6.6 &    2.763 & HMXB Be/X RXJ0059.2-7138; ASCA: 10$^{38}$ erg/s; XMM UL                              & 1 ps      po  xv  xs  oi  em  oo  \\ 
  5& 00 52 17.0 & -72 19 51 &111.0 & X &      &               &       &       &        &        &       &       &    4.780 & HMXB Be/X? XTEJ0052-723; [MA93]537 Halpha EQW=-43.3A OR AzV 129 Porb=23.9d       & 1 ps  px  po          oi: em      \\ 
  6& 00 57 02.3 & -72 25 55 &  0.5 & C &      &         14.77 & 15.76 & 15.96 & 15.75  &  -0.85 &  0.20 &   0.5 &    5.050 & HMXB Be/X IGR J00569-7226 (CBM13,CBK13,CBB15)                                    & 1 ps  px: po  xv  xs  oi  em      \\ 
  7& 01 02 53.4 & -72 44 35 &  0.5 & N &      &         13.68 & 14.79 & 14.98 & 15.08  &  -0.98 &  0.04 &   0.1 &    6.850 & HMXB Be/X XTEJ0103-728 = XMMUJ010253.1-724433 (HPK07)                            & 1 ps  px  po  xv  xs  oi  em  os  \\ 
  8& 00 54 10.0 & -72 25 44 &204.0 & I &      &               &       &       &        &        &       &       &    6.878 & HMXB? Integral 2.6sigma period (MCB07)                                           & 1 ps:                             \\ 
  9& 00 52 05.7 & -72 26 04 &  0.6 & C &  531 &         13.94 & 14.91 & 14.91 & 14.83  &  -0.97 &  0.05 &   0.1 &    7.780 & HMXB Be/X SMC X-3 (ECG04)                                                        & 1 ps  px  po  xv  xs  oi  em  oxo \\ 
 10& 00 57 58.4 & -72 22 30 &  1.5 & C &      &         14.60 & 15.64 & 15.72 & 15.93  &  -0.98 &  0.05 &   1.2 &    7.920 & HMXB Be/X SXP7.92 = CXOU J005758.4-722229 (CCM08,IED13,SC13)                     & 1 ps      po  xv  xs  oi  em      \\ 
 11& 00 51 53.2 & -72 31 48 &  1.0 & C &  506 &         13.66 & 14.79 & 14.38 & 14.40  &  -1.42 &  0.11 &   0.4 &    8.900 & HMXB Be/X RXJ0051.8-7231 (ISA97,SCB99)                                           & 1 ps  px  po          oi  em  os  \\ 
 12& 00 49 18.5 & -73 12 01 & 40.0 & A &      &               &       &       &        &        &       &       &    9.130 & HMXB Be/X AXJ0049-732 (YIT03,GCC08)                                              & 1 ps  px          xs:             \\ 
 13& 01 04 42.3 & -72 54 04 &  0.6 & C &      &         13.84 & 14.89 & 14.86 & 15.50  &  -1.06 &  0.06 &   0.8 &   11.480 & HMXB Be/X IGRJ01054-7253 = CXOU J010442.29-725404.4 (HEASARC cxoxassist)         & 1 ps  px  po  xv  xs  oi  em  os  \\ 
 14& 01 57 16.0 & -72 58 33 &  3.8 & S &      &               & 14.40 & 15.00 &        &        &       &   0.9 &   11.580 & HMXB Be/X? IGR J015712-7259 in Magellanic Bridge; opt. from NOMAD                & 1 ps  px  po  xv  xs  oi          \\ 
 15& 00 48 14.0 & -73 22 04 &  0.5 & N &      &         13.96 & 14.90 & 15.02 & 15.25  &  -0.85 &  0.05 &   0.4 &   11.866 & HMXB Be/X XMMUJ004813.9-732203 (SHC11)                                           & 1 ps          xv  xs  oi  em      \\ 
 16& 00 52 14.0 & -73 19 18 &  1.0 & C &  552 &         13.68 & 14.63 & 14.49 & 14.41  &  -1.06 &  0.09 &   0.6 &   15.300 & HMXB Be/X RXJ0052.1-7319 (LPM99,ISC99,HEP08)                                     & 1 ps  px  po          oi  em  oo  \\ 
 17& 00 50 44.6 & -73 16 05 &1800. & X &      &               &       &       &        &        &       &       &   16.600 & HMXB XTEJ0050-731 (LMP02) is NOT RXJ0051.9-7311; coord: 1deg FWHM                & 1 ps  px                      ox  \\ 
 18& 00 49 11.5 & -72 49 36 &  0.5 & N &      &         15.22 & 16.01 & 15.96 & 15.89  &  -0.82 &  0.04 &   1.1 &   18.370 & HMXB Be/X XTE J0055-727 = XMMU J004911.4-724939, XMM survey                      & 1 ps  px  po      xs  oi      os  \\ 
 19& 01 17 40.4 & -73 30 51 &  1.3 & N & 1845 &         13.30 & 14.14 & 14.18 & 14.09  &  -0.82 &  0.06 &   1.3 &   22.070 & HMXB Be/X RXJ0117.6-7330 (MFH99,CRW97) weak in XMM survey                        & 1 ps          xv      oi  em  oo  \\ 
 20& 00 48 14.2 & -73 10 04 &  0.6 & N &      &         15.06 & 15.56 & 15.30 & 15.51  &  -0.69 &  0.06 &   0.1 &   25.550 & HMXB Be/X XMMU J004814.1-731003 =? RXTE 25.5s or 51s pulsar (LMP02,GCC08)        & 1 ps              xs  oi  em  oo  \\ 
 21& 01 11 08.6 & -73 16 46 &  0.7 & N &      &         14.40 & 15.42 & 15.52 & 15.29  &  -0.96 &  0.05 &   0.1 &   31.030 & HMXB Be/X XTEJ0111.2-7317 (CLC98,ISC99); in XMM survey (SHP13)                   & 1 ps      po          oi  em  oo  \\ 
 22& 00 53 55.2 & -72 26 46 &  0.6 & C &      &         13.63 & 14.65 & 14.72 & 14.58  &  -0.97 &  0.05 &   0.9 &   46.630 & HMXB Be/X XTEJ0053-724 (CML98), Chandra (MCS08), Swift (SPH10)                   & 1 ps  px  po  xv  xs  oi  em  oxo \\ 
 23& 00 54 56.3 & -72 26 47 &  0.6 & N &  810 &         14.15 & 15.21 & 15.27 & 15.11  &  -1.03 &  0.04 &   0.1 &   59.070 & HMXB Be/X RXJ0054.9-7226=XTEJ0055-724 (MLS98,SCI98,SCB99,SPH03)                  & 1 ps  px  po          oi  em  oo  \\ 
 24& 01 07 12.6 & -72 35 34 &  0.9 & C & 1619 &         14.72 & 15.55 & 15.64 & 15.73  &  -0.77 &  0.09 &   0.5 &   65.780 & HMXB Be/X CXOUJ010712.6-723533=2E0105.7-7251=RXJ0107.1-7235=AXJ0107.2-7234 MCS07 & 1 ps      po      xs  oi  em      \\ 
 25& 00 49 03.3 & -72 50 52 &  0.6 & N &      &         16.02 & 16.86 & 16.78 & 16.61  &  -0.90 &  0.06 &   0.4 &   74.670 & HMXB Be/X AXJ0049-729 (CML98,SCB99,YIT03), XMM survey (SHP13)                    & 1 ps  px  po  xv  xs  oi  em  os  \\ 
 26& 00 52 08.9 & -72 38 03 &  0.6 & C &      &         14.33 & 15.15 & 15.23 & 14.83  &  -0.76 &  0.05 &   0.8 &   82.400 & HMXB Be/X XTEJ0052-725 (CMM02,ECC03,GCC08)                                       & 1 ps  px              oi  em  ox  \\ 
 27& 00 50 57.1 & -72 13 33 &  0.5 & N &  413 &         14.01 & 14.99 & 15.06 & 14.86  &  -0.93 &  0.06 &   0.7 &   91.120 & HMXB Be/X AXJ0051-722 (CML98,SCB99,CBB11)                                        & 1 ps  px  po  xv      oi  em  oo  \\ 
 28& 00 53 53.0 & -72 26 42 &1800. & X &      &               &       &       &        &        &       &       &   95.000 & HMXB XTE SMC95 (LCP02), Porb? (GCC08), position? 1deg FWHM                       & 1 ps                              \\ 
 29& 00 57 27.1 & -73 25 19 &  0.8 & C &      &         14.73 & 15.62 & 15.67 & 15.61  &  -0.86 &  0.12 &   0.4 &  101.160 & HMXB Be/X AXJ0057.4-7325=RXJ0057.3-7325 (YTK00,MCS07), Porb? (GCC08)             & 1 ps  px  po      xs  oi  em: oo  \\ 
 30& 00 53 24.0 & -72 27 16 &  1.0 & C &  667 &         15.06 & 16.11 & 16.19 & 16.14  &  -0.99 &  0.09 &   0.7 &  138.000 & HMXB Be/X CXOU J005323.8-722715 (ECG04,MCN08)                                    & 1 ps  px  po      xs  oi  em  ox  \\ 
 31& 00 56 05.7 & -72 21 59 &  0.9 & N &  904 &         14.75 & 15.84 & 15.88 & 15.98  &  -1.06 &  0.04 &   0.8 &  140.100 & HMXB Be/X XMMUJ005605.2-722200 2E0054.4-7237 (HS00,SPH03,CEG05)                  & 1 ps      po          oi  em  oo: \\ 
 32& 00 53 53.0 & -72 26 42 &1800. & X &      &               &       &       &        &        &       &       &  144.100 & HMXB XTE SMC144s (CMM03,GCC08), position? 1deg FWHM                              & 1 ps                              \\ 
 33& 00 57 50.3 & -72 07 57 &  1.0 & C & 1038 &         14.63 & 15.64 & 15.69 & 15.48  &  -0.97 &  0.14 &   1.0 &  152.340 & HMXB Be/X RX J0057.8-7207 (HS00), CXOU J005750.3-720756 (SPH03,MFL03,CEG05)      & 1 ps          xv  xs  oi  em      \\ 
 34& 01 07 43.3 & -71 59 54 &  0.6 & N & 1640 &         15.47 & 15.99 & 16.27 & 16.47  &  -0.32 &  0.05 &   0.9 &  153.990 & HMXB Be/X XMMUJ010743.1-715953, XMM survey, Pspin 2sigma (CHS12)                 & 1 ps:             xs  oi  em      \\ 
 35& 00 52 55.1 & -71 58 06 &  0.5 & N &  623 &         14.37 & 15.48 & 15.53 & 15.34  &  -1.07 &  0.06 &   0.6 &  169.300 & HMXB Be/X RXJ0052.9-7158=XTEJ0054-720=AXJ0052.9-7157=XMMUJ005255.0-715808 CSM97  & 1 ps  px  po  xv      oi  em      \\ 
 36& 00 51 51.9 & -73 10 33 &  0.9 & C &  504 &         13.39 & 14.38 & 14.45 & 14.28  &  -0.94 &  0.05 &   1.2 &  172.000 & HMXB Be/X RXJ0051.9-7311=AXJ0051.6-7311 (SCC99,YTI00) Porb? (SCM11)              & 1 ps  px  po      xs  oi  em  oo  \\ 
 37& 01 01 52.3 & -72 23 33 &  0.5 & N & 1288 &         13.84 & 14.94 & 14.94 & 14.85  &  -1.09 &  0.20 &   0.4 &  175.400 & HMXB Be/X RXJ0101.8-7223=AXJ0101.8-7223=XMMUJ010152.4-722336 (HS00,YIT03,TDC11)  & 1 ps              xs  oi  em      \\ 
 38& 00 59 21.0 & -72 23 17 &  0.5 & N &      &         13.81 & 14.96 & 14.98 & 15.07  &  -1.15 &  0.04 &   0.5 &  201.900 & HMXB Be/X RXJ0059.3-7223 (KPF99) XMMUJ005921.0-722317 (SPH03,MLM04) Porb?(GCC08) & 1 ps  px  po      xs  oi      oo  \\ 
 39& 00 59 28.9 & -72 37 04 &  0.5 & N & 1147 &         14.57 & 15.58 & 15.53 & 15.34  &  -1.04 &  0.06 &   0.8 &  202.520 & HMXB Be/X XMMU J005929.0-723703 (HEP08,SCM11), XMM survey (SHP13)                & 1 ps      po  xv  xs  oi  em  oo  \\ 
 40& 00 50 11.1 & -73 00 25 &  0.5 & N &      &         14.03 & 14.99 & 15.02 & 15.12  &  -0.94 &  0.04 &   1.1 &  214.030 & HMXB Be/X XMMUJ005011.1-730026 (XMM survey: CHS11), Porb (SCU13)                 & 1 ps      po  xv  xs  oi  em: oo  \\ 
 41& 00 47 23.3 & -73 12 28 &  0.5 & N &  172 &         15.14 & 16.11 & 16.03 & 15.99  &  -1.03 &  0.06 &   0.8 &  263.000 & HMXB Be/X RXJ0047.3-7312 (HS00) XMMUJ004723.7-731226 (HP04) (CEG05,SC05)         & 1 ps      po  xv  xs  oi  em  oo  \\ 
 42& 01 32 51.4 & -74 25 45 &  0.4 & N &      &               & 14.90 & 15.36 &        &        &       &   0.4 &  264.520 & HMXB Be/X XMMSL1 J013250.6-742544 = Swift J0132.5-7425 in Wing/Bridge (SHV14)    & 1 ps              xs  oi          \\ 
 43& 00 57 49.4 & -72 02 36 &  0.5 & N & 1036 &         14.53 & 15.54 & 15.65 & 15.52  &  -0.92 &  0.04 &   1.1 &  280.400 & HMXB Be/X RXJ0057.8-7202=AXJ0058-72.0 (TIY99,HS00,SPH03,CEG05)                   & 1 ps  px: po  xv  xs  oi  em  ox  \\ 
 44& 00 58 12.7 & -72 30 48 &  0.5 & N &      &         13.91 & 14.98 & 14.87 & 14.62  &  -1.14 &  0.07 &   0.5 &  291.330 & HMXB Be/X RXJ0058.2-7231 (SCC99,SPH03,EC03) = XTEJ0051-727 (HEP08)               & 1 ps  px  po      xs  oi  em  oo  \\ 
 45& 00 50 48.1 & -73 18 18 &  0.9 & N &  396 &         14.20 & 15.19 & 15.07 & 15.00  &  -1.08 &  0.06 &   0.3 &  292.700 & HMXB Be/X (SG05), CXOU J005047.9-731817 (LZH10,SHP13,EIS13)                      & 1 ps          xv  xs  oi  em      \\ 
 46& 01 01 02.8 & -72 06 58 &  1.2 & C & 1240 &         14.67 & 15.71 & 15.79 & 15.62  &  -0.99 &  0.05 &   0.7 &  304.500 & HMXB Be/X RXJ0101.0-7206 (KP96,SCB99,MFL03)                                      & 1 ps      po  xv  xs  oi  em  oo  \\ 
 47& 00 50 44.6 & -73 16 05 &  0.6 & C &  387 &         14.46 & 15.37 & 15.48 & 15.31  &  -0.83 &  0.06 &   0.7 &  323.200 & HMXB Be/X RXJ0050.8-7316=AXJ0051-733 (SCC99,CHL02,YIT03,GCC08)                   & 1 ps  px          xs  oi  em  ox  \\ 
 48& 00 52 52.1 & -72 17 15 &  0.6 & N &      &         15.83 & 16.41 & 16.62 & 16.82  &  -0.43 &  0.05 &   0.5 &  325.400 & HMXB Be/X XMMU J005252.1-721715 (HEP08) = CXOU J005252.2-721715 (CSC08,LZH10)    & 1 ps      po      xs  oi      oo  \\ 
 49& 00 54 03.9 & -72 26 32 &  0.6 & N &      &         13.82 & 14.95 & 14.94 & 14.79  &  -1.13 &  0.07 &   0.4 &  341.900 & HMXB Be/X XMMU J005403.8-722632 (HEP08) = CXOU J005403.9-722633 (LZH10)          & 1 ps              xs  oi          \\ 
 50& 01 03 13.9 & -72 09 14 &  0.9 & C & 1367 &         13.65 & 14.78 & 14.84 & 14.69  &  -1.08 &  0.05 &   0.6 &  345.200 & HMXB Be/X SAXJ0103.2-7209 (HS94,ISC98,CO00,ICC00) = AX J0103-722 (YK98)          & 1 ps      po  xv  xs  oi  em  oo  \\ 
 51& 01 01 20.7 & -72 11 19 &  0.5 & N & 1257 &         14.41 & 15.36 & 15.55 & 15.53  &  -0.82 &  0.06 &   0.4 &  455.000 & HMXB Be/X RXJ0101.3-7211 (SHK01,SCL04,CEG05)                                     & 1 ps      po      xs  oi  em  oo  \\ 
 52& 00 54 55.9 & -72 45 11 &  0.5 & N &  809 &         14.04 & 14.97 & 15.00 & 14.90  &  -0.91 &  0.18 &   0.3 &  499.200 & HMXB Be/X CXOUJ005455.6-724510 (ECG04) = XMMUJ005455.4-724512 (HPS04) (SC05)     & 1 ps  px  po      xs  oi  em  oxo \\ 
 53& 01 02 47.5 & -72 04 51 &  0.8 & N &      &         14.69 & 15.74 & 16.04 & 16.31  &  -0.84 &  0.05 &   0.5 &  522.500 & HMXB Be/X 2XMM J010247.4-720449 = Suzaku J0102-7204 (SHP11,WTE12,HST12,WTE13)    & 1 ps          xv  xs  oi  em      \\ 
 54& 00 57 36.2 & -72 19 34 &  0.7 & C & 1020 &         14.95 & 16.00 & 15.99 & 15.80  &  -1.06 &  0.21 &   0.7 &  565.000 & HMXB Be/X CXOUJ005736.2-721934 (MFL03,SPH03,SCL04,CEG05)                         & 1 ps  px  po  xv  xs: oi  em  oxo \\ 
 55& 00 55 35.4 & -72 29 07 &  0.5 & N &      &         13.60 & 14.65 & 14.69 & 14.67  &  -1.02 &  0.04 &   1.0 &  644.600 & HMXB Be/X XMMU J005535.2-722906 (HEP08)                                          & 1 ps              xs  oi          \\ 
 56& 00 55 18.3 & -72 38 52 &  0.5 & N &      &         15.45 & 15.92 & 15.77 & 15.65  &  -0.58 &  0.08 &   1.0 &  701.600 & HMXB Be/X XMMU J005517.9-723853 (HPS04,RCU11), Vmag from ZHT02 wrong             & 1 ps      po:     xs  oi      oo  \\ 
 57& 01 05 55.4 & -72 03 49 &  0.5 & N & 1557 &         14.50 & 15.64 & 15.70 & 15.57  &  -1.09 &  0.39 &   1.3 &  726.000 & HMXB Be/X RXJ0105.9-7203=AXJ0105.8-7203 (HS00,YIT03,SPH03,EH08)                  & 1 ps          xv  xs  oi  em      \\ 
 58& 00 49 42.1 & -73 23 15 &  0.5 & N &  315 &         14.09 & 15.00 & 14.85 & 14.70  &  -1.01 &  0.20 &   0.4 &  755.500 & HMXB Be/X RXJ0049.7-7323=AXJ0049.5-7323 (YIU00,HS00,EC03,HP04,SCL04)             & 1 ps  px  po  xv  xs  oi  em  oxo \\ 
 59& 00 49 29.7 & -73 10 59 &  0.6 & N &  300 &         15.46 & 16.35 & 16.15 & 16.00  &  -1.03 &  0.06 &   0.9 &  894.000 & HMXB Be/X RXJ0049.5-7310 = CXO J004929.7-731058 (HS00,HP04,LZH10)                & 1 ps      po  xv  xs  oi  em      \\ 
 60& 01 02 06.7 & -71 41 16 &  0.8 & C & 1301 &         13.53 & 14.58 & 14.37 & 14.42  &  -1.19 &  0.27 &   0.3 &  967.000 & HMXB Be/X CXOU J010206.6-714115 (MCS07,SCM07,HEP08) Porb? (SCU09)                & 1 ps      po      xs  oi  em  oo  \\ 
 61& 01 27 46.0 & -73 32 56 &  0.6 & C &      &         13.36 & 14.32 & 14.36 &        &  -0.93 &       &   0.1 & 1062.000 & HMXB Be/X CXO J012745.97-733256.5 (HSF12,HOG12,SHO13) opt. from M02              & 1 ps              xs  oi  em      \\ 
 62& 01 03 37.5 & -72 01 33 &  1.1 & C & 1393 &         13.47 & 14.55 & 14.65 & 14.69  &  -1.01 &  0.04 &   0.1 & 1323.000 & HMXB Be/X RXJ0103.6-7201 (HS00,HP05,SC06)                                        & 1 ps      po: xv  xs  oi  em      \\ 
 63& 00 54 46.2 & -72 25 23 &  1.0 & C &  798 &         14.39 & 15.50 & 15.36 & 15.25  &  -1.21 &  0.13 &   0.8 & 4693.000 & HMXB Be/X CXOU J005446.2-722523 (AHZ09,LZH10) XMMU J005446.3-722523 (SHP13)      & 1 ps:         xv  xs  oi  em      \\ 
 64& 00 32 56.2 & -73 48 20 & 12.9 & R &      &         16.18 & 16.72 & 16.90 & 16.86  &  -0.41 &  0.18 &  13.5 &          & HMXB Be/X RXJ0032.9-7348 (KP96), 2 Be candidates (SCB99), no XMM coverage        & 4                     oi: em      \\ 
 65& 00 42 07.8 & -73 45 03 &  0.7 & N &      &         16.19 & 16.68 & 16.78 & 16.90  &  -0.42 &  0.10 &   1.4 &          & HMXB Be/X XMMU J004207.7-734503(gam=0.47-0.96)=?AX J0042.0-7344(0.9-2.9) (SHP13) & 2                 xs  oi          \\ 
 66& 00 43 15.9 & -73 24 39 &  1.5 & N &      &         15.91 & 16.68 & 16.74 & 16.84  &  -0.73 &  0.09 &   2.7 &          & HMXB unlikely, XMMU J004315.8-732439 (weak source SHP13), no Balmer em. (Met16)  & 6                     oi:         \\ 
 67& 00 45 00.2 & -73 42 47 &  1.7 & N &      &         15.28 & 15.58 & 15.57 & 15.54  &  -0.30 &  0.06 &   1.5 &          & HMXB? Be/X? XMMU J004500.2-734246 (SHP13)                                        & 6                     oi:         \\ 
 68& 00 45 38.0 & -73 13 54 & 29.4 & R &      &         12.16 & 12.97 & 13.02 & 13.02  &  -0.78 &  0.02 &  19.9 &          & HMXB Be/X? RXJ0045.6-7313, [MA93]114 or AzV9? (HS00), no XMM detection (SHP13)   & 6                     oi:         \\ 
 69& 00 48 18.7 & -73 21 00 &  0.6 & N &      &         15.66 & 16.43 & 16.18 & 15.79  &  -0.95 &  0.05 &   0.2 &          & HMXB rejected, XMMU J004818.6-732059 (SG05,AZH09), QSO (KK09,SHP13,SDF13)        & 6                 xs: oi:         \\ 
 70& 00 48 34.1 & -73 02 31 &  0.7 & N &  238 &         13.97 & 14.78 & 14.78 & 14.76  &  -0.81 &  0.04 &   0.5 &          & HMXB Be/X RX J0048.5-7302 = XMMU J004834.5-730230 (HS00,SG05,HEP08,KBS14)        & 2         po  xv  xs  oi  em      \\ 
 71& 00 48 49.0 & -73 16 25 &  4.4 & C &  258 &         13.59 & 14.50 & 14.56 & 14.28  &  -0.87 &  0.06 &   1.8 &          & HMXB? Be/X? weak Chandra source (LHZ10), not in EPG10, source real?              & 4                     oi: em      \\ 
 72& 00 48 55.6 & -73 49 46 &  0.6 & N &      &         13.95 & 14.69 & 14.93 & 14.94  &  -0.56 &  0.06 &   1.3 &          & HMXB Be/X XMMU J004855.5-734946, bright in XMM survey, gamma=0.8 (SHP13)         & 2                 xs  oi:         \\ 
 73& 00 49 02.7 & -73 27 07 &  1.6 & N &      &         14.74 & 15.61 & 15.79 & 15.97  &  -0.75 &  0.06 &   3.5 &          & HMXB rejected, XMMU J004902.6-732707 (SHP13), no Balmer emission (Met16)         & 6                     oi: nem     \\ 
 74& 00 49 13.6 & -73 11 38 &  0.5 & N &      &         15.71 & 16.63 & 16.44 & 16.28  &  -1.06 &  0.06 &   0.3 &          & HMXB Be/X RXJ0049.2-7311 (FPH00,SG05,CEG05,HEP08,RCU11) =?SXP9.13=AXJ0049-732    & 2                 xs  oi  em  oo  \\ 
 75& 00 49 22.2 & -73 20 06 &  3.4 & C &      &         16.08 & 16.71 & 16.62 & 16.85  &  -0.70 &  0.07 &   1.7 &          & HMXB? weak Chandra source, blue early-type star (LZH10), not in EPG10, real?     & 6                     oi:         \\ 
 76& 00 49 30.6 & -73 31 09 &  0.6 & N &  302 &         13.73 & 14.60 & 14.64 & 14.55  &  -0.83 &  0.06 &   0.4 &          & HMXB Be/X RX J0049.5-7331 = XMMU J004930.6-733109 (HS00,HEP08,SHP13)             & 2                 xs  oi  em      \\ 
 77& 00 49 41.7 & -72 48 43 &  0.6 & C &      &         15.06 & 16.24 & 15.99 & 15.62  &  -1.35 &  0.15 &   0.7 &          & HMXB? CXOU J004941.43-724843.8 (AZH09,MZA13), not in EPG10, source real?         & 6                     oi: em:     \\ 
 78& 00 50 04.4 & -73 14 26 &  1.6 & C &      &         14.72 & 15.56 & 15.50 & 15.72  &  -0.88 &  0.06 &   0.9 &          & HMXB? weak Chandra source, blue early-type star (LZH10), not in EPG10, real?     & 6                     oi:         \\ 
 79& 00 50 12.2 & -73 11 56 &  1.7 & C &  341 &         14.56 & 15.47 & 15.31 & 14.98  &  -1.02 &  0.09 &   1.2 &          & HMXB? Be/X? weak Chandra source (LZH10)                                          & 4                     oi: em      \\ 
 80& 00 50 35.5 & -73 14 01 &  1.1 & C &      &         15.21 & 16.10 & 15.99 & 15.99  &  -0.97 &  0.05 &   1.5 &          & HMXB? weak Chandra source, blue early-type star (LZH10)                          & 6                     oi:         \\ 
 81& 00 50 36.0 & -73 17 39 &  0.9 & C &  374 &         14.72 & 15.69 & 15.61 & 15.40  &  -1.03 &  0.05 &   0.8 &          & HMXB? Be/X? weak Chandra source (LZH10)                                          & 4                     oi: em      \\ 
 82& 00 50 46.9 & -73 32 48 & 33.7 & R &      &         14.96 & 15.64 & 15.61 & 15.42  &  -0.70 &  0.06 &   9.9 &          & HMXB? Be/X? RX J0050.7-7332 [MA93]393? (HS00) XMM source 11arcsec away AGN?      & 6                     oi:         \\ 
 83& 00 50 47.8 & -73 17 36 &  1.0 & C &      &         15.65 & 16.58 & 16.58 & 16.47  &  -0.94 &  0.05 &   0.5 &          & HMXB? weak Chandra source, blue early-type star (LZH10)                          & 6                     oi:         \\ 
 84& 00 50 57.3 & -73 10 08 &  0.5 & N &  414 &         13.55 & 14.43 & 14.35 & 14.40  &  -0.94 &  0.06 &   0.6 &          & HMXB Be/X RXJ0050.9-7310 = CXO J005057.2-731008 (HS00,SG05,LZH10,SHP13)          & 2                 xs  oi  em      \\ 
 85& 00 51 05.7 & -73 13 12 &  1.2 & C &      &         14.84 & 15.77 & 15.70 & 15.52  &  -0.99 &  0.06 &   0.7 &          & HMXB? weak Chandra source, blue early-type star (LZH10)                          & 5                     oi          \\ 
 86& 00 51 17.0 & -73 16 06 &  1.0 & C &  448 &         14.13 & 15.19 & 15.00 & 14.92  &  -1.19 &  0.05 &   0.9 &          & HMXB? Be/X? weak Chandra source (LZH10)                                          & 4                     oi: em      \\ 
 87& 00 51 19.6 & -72 50 44 & 15.6 & R &      &         15.45 & 16.36 & 16.31 & 16.01  &  -0.95 &  0.05 &  14.6 &          & HMXB? Be/X? RXJ0051.3-7250 [MA93]447? (HS00) XMM source 17.7arcsec away, AGN?    & 6                     oi:         \\ 
 88& 00 51 33.3 & -73 30 12 &  1.5 & N &      &         15.85 & 16.65 & 16.58 & 16.72  &  -0.85 &  0.32 &   4.4 &          & HMXB rejected, XMMU J005133.2-733012 (SHP13), no Balmer emission (Met16)         & 6                     oi: nem     \\ 
 89& 00 51 46.1 & -73 07 04 &  1.1 & N &      &         16.05 & 16.71 & 16.73 & 16.90  &  -0.64 &  0.05 &   2.9 &          & HMXB rejected, XMMU J005146.1-730704 (SHP13), no Balmer emission (Met16)         & 6                     oi: nem     \\ 
 90& 00 51 54.2 & -72 55 36 & 40.0 & E &      &         14.43 & 15.49 & 15.40 & 15.26  &  -1.13 &  0.08 &  28.2 &          & HMXB Be/X? RXJ0051.9-7255 [MA93]521? (HS00) no XMM detection                     & 6                     oi:         \\ 
 91& 00 51 59.6 & -73 29 26 &  3.0 & S &      &         14.41 & 15.23 & 15.18 & 14.98  &  -0.85 &  0.06 &   4.7 &          & HMXB Be/X IGR J00515-7328 (CBB10,SHP11,K11)                                      & 2             xv      oi:         \\ 
 92& 00 52 07.8 & -72 21 26 &  2.0 & N &      &         14.50 & 15.12 & 15.20 & 14.73  &  -0.57 &  0.04 &   1.4 &          & HMXB Be/X? XMMU J005207.8-722125 (LZH10,SHP13, unclassified in AZH09); SXP4.78?  & 6                     oi:         \\ 
 93& 00 52 15.4 & -73 19 15 &  1.0 & C &      &         14.81 & 15.76 & 15.90 & 16.10  &  -0.85 &  0.05 &   0.4 &          & HMXB Be/X? CXOUJ005215.4-731915 very close to SXP15.3 (LZH10,SHP13)              & 2                 xs  oi          \\ 
 94& 00 52 35.3 & -72 25 21 &  1.6 & N &      &         13.81 & 14.73 & 14.90 & 15.14  &  -0.80 &  0.04 &   5.7 &          & HMXB rejected, XMMU J005235.2-722520 (SHP13), Chandra position incompatible      & 6                     oi:         \\ 
 95& 00 52 37.3 & -72 27 32 &  1.2 & C &  590 &         13.88 & 14.99 & 14.98 & 14.76  &  -1.12 &  0.04 &   0.4 &          & HMXB? Be/X? weak Chandra source (LZH10)                                          & 4                     oi: em      \\ 
 96& 00 52 45.0 & -72 28 44 &  1.0 & C &      &         13.95 & 14.92 & 14.92 & 14.87  &  -0.97 &  0.07 &   0.4 &          & HMXB Be/X CXOUJ005245.0-722844 (LZH10)                                           & 3                     oi  em      \\ 
 97& 00 52 52.2 & -72 48 30 &  1.0 & C &  618 &         13.35 & 14.32 & 14.36 & 14.24  &  -0.93 &  0.05 &   0.6 &          & HMXB? peculiar CXOU J005252.2-724830 =?2E0051.1-7304 AzV138 (AZH09)              & 3                     oi  em      \\ 
 98& 00 52 59.5 & -72 54 02 &  2.1 & N &      &         16.52 & 16.98 & 16.79 & 16.43  &  -0.60 &  0.09 &   1.6 &          & HMXB? Be/X? XMMU J005259.4-725402; weak source in XMM survey (SHP13)             & 6                     oi:         \\ 
 99& 00 53 14.8 & -72 18 48 &  1.7 & N &      &         16.99 & 17.00 & 16.39 & 15.53  &  -0.45 &  0.05 &   2.7 &          & HMXB rejected, weak source, Chandra position incompatible (LZH10,SHP13)          & 6                     oi:         \\ 
100& 00 53 18.5 & -72 16 18 &  1.6 & N &      &         15.62 & 16.42 & 16.58 &        &  -0.68 &  0.05 &   2.3 &          & HMXB? Be/X? XMMU J005318.5-721617 (SHP13)                                        & 6                     oi:         \\ 
101& 00 53 29.2 & -72 33 48 &  2.7 & C &  677 &         13.69 & 14.63 & 14.62 & 14.57  &  -0.95 &  0.29 &   0.7 &          & HMXB? Be/X? weak Chandra source (LZH10)                                          & 4                     oi: em      \\ 
102& 00 53 31.8 & -72 18 45 &  5.2 & C &      &         15.08 & 15.86 & 16.03 & 16.25  &  -0.65 &  0.05 &   5.4 &          & HMXB unlikely, weak Chandra source (LZH10), not in EPG10, source real?           & 6                     oi: nem     \\ 
103& 00 53 34.6 & -72 08 42 &  0.9 & N &      &         15.87 & 16.19 & 16.19 & 16.16  &  -0.31 &  0.12 &   2.7 &          & HMXB unlikely, XMMU J005334.6-720842 (SHP13)                                     & 6                     oi: nem     \\ 
104& 00 53 41.8 & -72 53 10 &  0.8 & N &      &         13.66 & 14.74 & 14.66 & 14.49  &  -1.14 &  0.19 &   2.2 &          & HMXB Be/X XMMU J005341.7-725310 (SHP13)                                          & 2             xv      oi:         \\ 
105& 00 53 52.5 & -72 26 39 &  0.9 & C &  717 &         13.52 & 13.92 & 13.67 & 13.13  &  -0.58 &  0.12 &   1.3 &          & HMXB Be/X CXOU J005352.5-722639 close to SXP46.6 (LZH10)                         & 3                     oi  em      \\ 
106& 00 54 08.7 & -72 32 08 &  1.4 & N &      &         16.59 & 16.86 & 16.93 & 16.96  &  -0.22 &  0.06 &   1.1 &          & HMXB? Be/X? weak source detected by Chandra and XMM (LZH10,SHP13)                & 6                     oi:         \\ 
107& 00 54 09.3 & -72 41 43 &  1.4 & N &  739 &         13.10 & 13.77 & 13.82 & 12.62  &  -0.64 &  0.15 &   1.3 &          & HMXB rejected, XMMU J005409.2-724143 (SHP13,AZH09), sgB0[e] S18 (CBC13,MZA14)    & 3                     oi  em      \\ 
108& 00 54 19.2 & -72 20 49 &  5.8 & C &      &         16.04 & 16.92 & 16.97 & 17.20  &  -0.84 &  0.08 &   1.9 &          & HMXB unlikely, weak Chandra source (LHZ10), not in EPG10, source real?           & 6                     oi: nem     \\ 
109& 00 54 26.0 & -71 58 24 &  0.9 & N &      &         15.61 & 16.42 & 16.56 & 16.77  &  -0.71 &  0.08 &   2.4 &          & HMXB rejected, XMMU J005425.9-715824 (SHP13), no Balmer emission (Met16)         & 6                     oi: nem     \\ 
110& 00 55 07.3 & -72 08 26 &  1.7 & N &      &         16.05 & 16.77 & 16.87 & 16.94  &  -0.66 &  0.19 &   3.9 &          & HMXB unlikely, XMMU J005507.2-720825 (SHP13)                                     & 6                     oi: nem     \\ 
111& 00 55 07.7 & -72 22 40 &  0.9 & N &      &         13.23 & 14.26 & 14.38 & 14.58  &  -0.93 &  0.06 &   0.8 &          & HMXB Be/X? XMMU J005507.7-722240 = CXOU J005507.7-722241 (SHP13,LZH10)           & 5                     oi          \\ 
112& 00 55 35.0 & -71 33 41 &  1.3 & N &      &         15.10 & 15.94 & 16.08 & 16.37  &  -0.74 &  0.05 &   4.7 &          & HMXB rejected, XMMU J005535.0-713340 (SHP13), no Balmer emission (Met16)         & 6                     oi: nem     \\ 
113& 00 55 49.8 & -72 51 27 &  1.5 & N &      &         15.86 & 16.48 & 16.49 & 16.65  &  -0.62 &  0.06 &   1.0 &          & HMXB rejected, (SHP13), OGLEII eclipsing binary (WUK04)                          & 6                     oi: nem     \\ 
114& 00 56 05.5 & -72 00 11 &  2.0 & N &      &         15.69 & 16.60 & 16.72 & 16.97  &  -0.84 &  0.05 &   1.3 &          & HMXB Be/X XMMU J005605.8-720012 (NLM11,SHP13)                                    & 3                     oi  em      \\ 
115& 00 56 13.9 & -72 30 00 &  1.0 & N &      &         13.49 & 14.53 & 14.52 & 14.39  &  -1.05 &  0.04 &   0.7 &          & HMXB Be/X XMMU J005613.8-722959 (SHP13)                                          & 3                     oi  em      \\ 
116& 00 56 14.6 & -72 37 56 &  0.8 & N &  922 &         13.43 & 14.71 & 14.58 & 14.21  &  -1.37 &  0.34 &   0.7 &          & HMXB Be/X XMMU J005614.6-723755 (SHP13)                                          & 3                     oi  em      \\ 
117& 00 56 18.9 & -72 28 03 &  0.7 & N &      &         14.58 & 14.93 & 15.32 & 15.56  &  -0.07 &  0.35 &   3.1 &          & HMXB? Be/X? XMMU J005618.8-722802, Be star: NGC 330:KWBBe 224 (KWB99,SG05,SHP13) & 4                 xs: oi: em      \\ 
118& 00 56 19.0 & -72 15 06 &  1.8 & N &      &         15.18 & 16.06 & 16.13 & 16.12  &  -0.84 &  0.06 &   5.2 &          & HMXB rejected, XMMU J005619.0-721506 (SHP13), Chandra position incompatible      & 6                     oi:         \\ 
119& 00 57 23.7 & -72 23 56 &  0.8 & N &      &         13.60 & 14.64 & 14.71 & 14.92  &  -0.98 &  0.05 &   1.7 &          & HMXB Be/X? XMMU J005723.4-722356 (SG05,SHP13,MZA14) rel. large dist to ZHT02     & 2             xv      oi: em:     \\ 
120& 00 57 59.5 & -71 56 37 & 19.2 & R &      &         14.36 & 14.98 & 14.94 & 14.84  &  -0.65 &  0.20 &  21.0 &          & HMXB Be/X? RXJ0057.9-7156 [MA93]1044? (HS00) no XMM detection                    & 6                     oi:         \\ 
121& 01 00 30.3 & -72 20 33 &  1.0 & N & 1208 &         13.57 & 14.59 & 14.64 & 14.54  &  -0.98 &  0.17 &   0.7 &          & HMXB Be/X XMMUJ010030.2-722035 (SPH03,SG05)                                      & 3                     oi  em      \\ 
122& 01 00 37.3 & -72 13 17 &  0.9 & N &      &         15.59 & 16.51 & 16.68 & 16.83  &  -0.80 &  0.05 &   2.4 &          & HMXB rejected, XMMU J010037.3-721317 (SG05), AGN? (SHP13), no Balmer em. (Met16) & 6                     oi:         \\ 
123& 01 00 55.8 & -72 23 20 &  1.0 & N &      &               & 15.49 & 15.61 &        &        &       &   1.1 &          & HMXB? Be/X? (SHP13), Bmag from ZHT02 wrong, using B-V from M02                   & 3                     oi  em      \\ 
124& 01 01 47.6 & -71 55 51 &  0.9 & N & 1284 &         13.26 & 14.40 & 14.47 & 14.30  &  -1.09 &  0.05 &   1.4 &          & HMXB SSS Be/WD? XMMUJ010147.5-715550 (SHP12)                                     & 3                     oi  em      \\ 
125& 01 01 55.8 & -72 32 37 &  0.6 & N &      &         12.72 & 13.93 & 14.02 & 14.01  &  -1.14 &  0.10 &   0.8 &          & HMXB Be/X, wrong ID with SXP7.92 (CSM09), counterpart AzV285 (SHP13), (RCU11)    & 2         po  xv  xs  oi      oo  \\ 
126& 01 01 55.9 & -72 10 28 &  0.9 & N &      &         13.93 & 14.85 & 15.06 & 15.31  &  -0.77 &  0.40 &   0.9 &          & HMXB unlikely, XMMU J010155.8-721027 (SHP13), no Balmer emission (Met16)         & 5                     oi          \\ 
127& 01 03 28.5 & -72 06 51 &  0.6 & N &      &         15.46 & 16.32 & 16.47 & 16.78  &  -0.75 &  0.06 &   1.9 &          & HMXB rejected, (SG05,SHP13), OGLEII eclipsing binary (WUK04)                     & 6                     oi: nem     \\ 
128& 01 03 31.7 & -73 01 44 &  1.0 & N &      &         14.09 & 15.17 & 15.41 & 15.65  &  -0.91 &  0.04 &   1.5 &          & HMXB unlikely, XMMU J010331.7-730144 (SHP13), no Balmer emission (Met16)         & 6                     oi:         \\ 
129& 01 03 33.6 & -72 04 17 &  1.6 & N &      &         14.98 & 15.97 & 16.08 & 16.38  &  -0.91 &  0.10 &   4.9 &          & HMXB rejected, XMMU J010333.6-720417 (SHP13), Chandra position incompatible      & 6                     oi: nem     \\ 
130& 01 03 38.0 & -72 02 15 &  1.6 & N &      &         15.32 & 16.12 & 16.31 & 16.52  &  -0.67 &  0.07 &   4.5 &          & HMXB rejected, XMMU J010338.0-720215 (SHP13), Chandra position incompatible      & 6                     oi: nem     \\ 
131& 01 03 55.1 & -72 49 53 &  1.5 & N &      &         16.33 & 16.96 & 16.25 &        &  -1.15 &  0.18 &   3.7 &          & HMXB? Be/X? XMMU J010355.0-724952, in NGC376 (SHP13)                             & 6                     oi:         \\ 
132& 01 04 29.4 & -72 31 37 &  1.3 & N &      &         14.53 & 15.61 & 15.79 & 16.02  &  -0.96 &  0.08 &   1.4 &          & HMXB Be/X XMMU J010429.4-723136 (SHP13,MSH13)                                    & 2         po: xv  xs: oi          \\ 
133& 01 04 35.5 & -72 21 47 &  0.8 & N & 1470 &         14.07 & 15.18 & 15.13 & 14.98  &  -1.15 &  0.04 &   0.6 &          & HMXB Be/X RX J0104.5-7221 = XMMU J010435.4-722147 (HS00,SHP13)                   & 2                 xs  oi  em      \\ 
134& 01 04 48.5 & -71 45 42 &  1.5 & N &      &         17.10 & 17.30 & 16.87 & 16.24  &  -0.51 &  0.06 &   4.3 &          & HMXB unlikely, XMMU J010448.5-714541 (SHP13), no Balmer emission (Met16)         & 6                     oi:         \\ 
135& 01 06 00.8 & -72 33 04 &  1.9 & N &      &         15.24 & 16.18 & 16.27 & 16.47  &  -0.88 &  0.04 &   2.0 &          & HMXB unlikely, XMMU J010600.7-723303; weak source in XMM survey (SHP13)          & 6                     oi: nem     \\ 
136& 01 06 33.0 & -73 15 43 &  0.6 & N & 1592 &         13.78 & 14.68 & 15.06 & 15.07  &  -0.63 &  0.06 &   0.4 &          & HMXB Be/X XMMU J010633.1-731543 (CHS12)                                          & 2                 xs  oi  em      \\ 
137& 01 07 44.5 & -72 27 42 &  0.6 & C & 1641 &         14.26 & 15.47 & 15.49 & 15.31  &  -1.21 &  0.04 &   1.5 &          & HMXB Be/X CXOU J010744.51-722741.7 = Swift J010745.0-722740 (MSH14)              & 2             xv      oi  em      \\ 
138& 01 08 20.2 & -72 13 47 &  0.7 & N &      &         13.83 & 14.55 & 14.67 & 14.77  &  -0.62 &  0.23 &   2.2 &          & HMXB rejected, XMM (SHP13), Chandra (EPG10), no Balmer em. (Met16), AGN?         & 6                     oi:         \\ 
139& 01 15 03.5 & -73 28 19 &  1.0 & N &      &         15.60 & 16.34 & 16.48 & 16.69  &  -0.65 &  0.07 &   1.6 &          & HMXB unlikely, (XMM survey) HR3/4 just outside of HMXB criteria of SHP13         & 6                     oi: nem     \\ 
140& 01 19 03.5 & -73 12 21 &  1.5 & N &      &         15.18 & 15.93 & 16.04 & 16.14  &  -0.68 &  0.06 &   5.6 &          & HMXB unlikely, only 1 of 2 detections compatible with B-star (SHP13); SXP2.16?   & 6                     oi: nem     \\ 
141& 01 19 38.9 & -73 30 11 &  0.7 & N & 1867 &         14.95 & 15.78 & 15.85 & 15.61  &  -0.79 &  0.12 &   0.3 &          & HMXB Be/X? RX J0119.6-7330 (HS00,SG05,SHP13)                                     & 3                 xs: oi  em      \\ 
142& 01 21 41.0 & -72 57 33 &  4.0 & S & 1888 &         13.17 & 14.23 & 14.28 &        &        &       &   3.2 &          & HMXB Be/X IGR J01217-7257 (CBM14) opt. from M02                                  & 3         po          oi  em      \\ 
143& 01 23 27.5 & -73 21 23 &  1.1 & N &      &         14.48 & 15.39 & 15.45 &        &  -0.87 &       &   1.4 &          & HMXB Be/X RX J0123.4-7321=XMM J012327.4-732123 (SHP13, SHPU13); opt. from M02    & 2         po  xv  xs: oi          \\ 
144& 02 04 49.0 & -73 15 27 &108.0 & I &      &               &       &       &        &        &       &       &          & HMXB? Be/X? IGR J02048-7315, Magellanic Bridge (MBC10)                           & 6                                 \\ 
145& 02 06 45.7 & -74 27 46 &  4.0 & S &      &               & 14.44 & 14.77 &        &        &       &   2.5 &          & HMXB Be/X? SWIFT J0208.4-7428, Magellanic Bridge (MBC10,SCU14), opt. from NOMAD  & 2             xv  xs  oi  em      \\ 
146& 02 09 37.2 & -74 27 12 & 30.0 & R &      &               & 14.16 & 14.53 &        &        &       &  10.7 &          & HMXB Be/X RXJ0209.6-7427 Mag. Bridge, var~20 in PSPC lc (KH05), opt. from NOMAD  & 2             xv  xs  oi  em      \\ 
147& 02 22 01.0 & -75 57 59 &108.0 & I &      &               &       &       &        &        &       &       &          & HMXB? Be/X? IGR J02220-7558, Magellanic Bridge (MBC10)                           & 6                                 \\ 
148& 03 14 23.0 & -74 04 23 &300.0 & I &      &               &       &       &        &        &       &       &          & HMXB? Be/X? IGR J03144-7404, Magellanic Bridge (MBC10)                           & 6                                 \\ 

\end{longtable}
\end{landscape}
\end{longtab}

\addtocounter{table}{-1}

\begin{table*}
\caption[]{Key references.}
\begin{center}
\begin{tabular}{ll|ll}
\hline\hline\noalign{\smallskip}
\multicolumn{1}{l}{Code\tablefootmark{a}} &
\multicolumn{1}{l}{Reference} &
\multicolumn{1}{l}{Code\tablefootmark{a}} &
\multicolumn{1}{l}{Reference} \\
\noalign{\smallskip}\hline\noalign{\smallskip}
AHZ09      & {\citet{2009ApJ...707.1080A}} &
AZH09      & {\citet{2009ApJ...697.1695A}} \\
BCS01      & {\citet{2001MNRAS.320..281B}} &
BLK10      & {\citet{2010AJ....140..416B}} \\
CBB10      & {\citet{2010MNRAS.406.2533C}} &
CBB11      & {\citet{2011ATel.3396....1C}} \\
CBB15      & {\citet{2015MNRAS.447.2387C}} &
CBC13      & {\citet{2013A&A...560A..10C}} \\
CBK13      & {\citet{2013ATel.5662....1C}} &
CBM13      & {\citet{2013ATel.5547....1C}} \\
CBM14      & {\citet{2014ATel.5806....1C}} &
CCM08      & {\citet{2008ATel.1600....1C}} \\
CEG05      & {\citet{2005MNRAS.356..502C}} &
CHL02      & {\citet{2002MNRAS.332..473C}} \\
CHS11      & {\citet{2011MNRAS.414.3281C}} &
CHS12      & {\citet{2012MNRAS.424..282C}} \\
CLC98      & {\citet{1998IAUC.7048....1C}} &
CML98      & {\citet{1998IAUC.6803....1C}} \\
CMM02      & {\citet{2002IAUC.7932....2C}} &
CMM03      & {\citet{2003ATel..163....1C}} \\
CNC01      & {\citet{2001A&A...374.1009C}} &
CO00       & {\citet{2000MNRAS.311..169C}} \\
CRW97      & {\citet{1997ApJ...474L.111C}} &
CSC08      & {\citet{1999AJ....117..927S}} \\
CSM97      & {\citet{1997PASP..109...21C}} &
CSM09      & {\citet{2009MNRAS.394.2191C}} \\
CTO98      & {\citet{1998IAUC.7062....1C}} &
EC03       & {\citet{2003MNRAS.338..428E}} \\
ECC03      & {\citet{2003ATel..215....1E}} &
ECG04      & {\citet{2004MNRAS.353.1286E}} \\
EH08       & {\citet{2008A&A...485..807E}} &
EHI04      & {\citet{2004MNRAS.353..601E}} \\
EIS13      & {\citet{2013MNRAS.433.3464E}} &
ELS06      & {\citet{2006ASPC..353...41E}} \\
EPG10      & {\citet{2010ApJS..189...37E}} &
FPH00      & {\citet{2000A&A...361..823F}} \\
GCC08      & {\citet{2008ApJS..177..189G}} &
HEP08      & {\citet{2008A&A...489..327H}} \\
HPK07      & {\citet{2007ATel.1095....1H}} &
HP05       & {\citet{2005A&A...438..211H}} \\
HPS04      & {\citet{2004A&A...420L..19H}} &
HOG12      & {\citet{2012MNRAS.420L..13H}} \\
HP04       & {\citet{2004A&A...414..667H}} &
HS94       & {\citet{1994AJ....107.1363H}} \\
HS00       & {\citet{2000A&A...359..573H}} &
HSF12      & {\citet{2012A&A...537L...1H}} \\
HST12      & {\citet{2012ATel.4648....1H}} &
ICC00      & {\citet{2000ApJ...531L.131I}} \\
IED13      & {\citet{2013ATel.5552....1I}} &
ISA97      & {\citet{1997ApJ...484L.141I}} \\
ISC98      & {\citet{1998IAUC.6999....1I}} &
ISC99      & {\citet{1999IAUC.7101....1I}} \\
K11        & {\citet{2011ATel.3578....1K}} &
KBS14      & {\citet{2014A&A...562A.125K}} \\
KH05       & {\citet{2005A&A...435....9K}} &
KPF99      & {\citet{1999A&AS..136...81K}} \\
KWB99      & {\citet{1999A&AS..134..489K}} &
LCP02      & {\citet{2002A&A...385..464L}} \\
LMP02      & {\citet{2002ApJ...567L.129L}} &
LPM99      & {\citet{1999IAUC.7081....4L}} \\
LZH10      & {\citet{2010ApJ...716.1217L}} &
MA93       & {\citet{1993A&AS..102..451M}} \\
M02        & {\citet{2002ApJS..141...81M}} &
MBC10      & {\citet{2010MNRAS.403..709M}} \\
MCB07      & {\citet{2007MNRAS.382..743M}} &
MCN08      & {\citet{2008MNRAS.388.1198M}} \\
MCS07      & {\citet{2007MNRAS.376..759M}} &
MCS08      & {\citet{2008MNRAS.384..821M}} \\
Met16      & McBride et al. in preparation &%
MFH99      & {\citet{1999ApJ...518L..99M}} \\
MFL03      & {\citet{2003ApJ...584L..79M}} &
MLM04      & {\citet{2004ApJ...609..133M}} \\
MLS98      & {\citet{1998IAUC.6818....1M}} &
MPP10      & {\citet{2010A&A...519A..96M}} \\
MSH13      & {\citet{2013ATel.5674....1M}} &
MSH14      & {\citet{2014ATel.5778....1M}} \\
MZA14      & {\citet{2014MNRAS.438.2005M}} &
NLM11      & {\citet{2011A&A...532A.153N}} \\
NHS03      & {\citet{2003ApJ...586..983N}} &
RCU11      & {\citet{2011MNRAS.413.1600R}} \\
SC05       & {\citet{2005AJ....130.2220S}} &
SC06       & {\citet{2006ATel..716....1S}} \\
SC13       & {\citet{2013ATel.5556....1S}} &
SCB99      & {\citet{1999MNRAS.309..421S}} \\
SCC99      & {\citet{1999AJ....117..927S}} &
SCI98      & {\citet{1998IAUC.6818R...1S}} \\
SCL04      & {\citet{2004AJ....127.3388S}} &
SCM07      & {\citet{2007MNRAS.381.1561S}} \\
SCM11      & {\citet{2011MNRAS.412..391S}} &
SCU09      & {\citet{2009ATel.1953....1S}} \\
SCU13      & {\citet{2013ATel.4936....1S}} &
SCU14      & {\citet{2014ATel.5776....1S}} \\
SDF13      & {\citet{2013A&A...558A.101S}} &
SG05       & {\citet{2005MNRAS.362..879S}} \\
SHC11      & {\citet{2011A&A...527A.131S}} &
SHK01      & {\citet{2001A&A...369L..29S}} \\
SHO13      & {\citet{2013A&A...556A.139S}} &
SHP11      & {\citet{2011ATel.3761....1S}} \\
SHP12      & {\citet{2012A&A...537A..76S}} &
SHP13      & {\citet{2013A&A...558A...3S}} \\
SHPU13     & {\citet{2013A&A...551A..96S}} &
SHV14      & {\citet{2014MNRAS.444.3571S}} \\
SPH03      & {\citet{2003A&A...403..901S}} &
SPH10      & {\citet{2010ATel.2876....1S}} \\
SHP11      & {\citet{2011ATel.3575....1S}} &
TDC11      & {\citet{2011ATel.3311....1T}} \\
TIY99      & {\citet{1999PASJ...51L..21T}} &
WTE12      & {\citet{2012ATel.4628....1W}} \\
WTE13      & {\citet{2013PASJ...65L...2W}} &
WUK04      & {\citet{2004AcA....54....1W}} \\
YIT03      & {\citet{2003PASJ...55..161Y}} &
YIU00      & {\citet{2000PASJ...52L..73Y}} \\
YK98       & {\citet{1998IAUC.7009....3Y}} &
YTI00      & {\citet{2000PASJ...52L..37Y}} \\
YTK00      & {\citet{2000PASJ...52L..53Y}} &
ZHT02      & {\citet{2002AJ....123..855Z}} \\
\noalign{\smallskip}\hline
\end{tabular}
\tablefoot{\tablefoottext{a}{As used in the comment column of Table~\ref{tab:smccat}.}}
\end{center}
\label{tab:keyref}
\end{table*}

\end{document}